\documentclass[iop,numberedappendix,twocolappendix]{emulateapj}

\newcommand{\beq}{\begin{equation}}
\newcommand{\eeq}{\end{equation}}
\newcommand{\beqa}{\begin{eqnarray}}
\newcommand{\eeqa}{\end{eqnarray}}

\newcommand{\NH}{N_{\rm{H}}}

\newcommand{\fint}{f_{\rm HX}}

\newcommand{\Lint}{L_{\rm HX}}
\newcommand{\Lintn}{L_{{\rm HX},43}}
\newcommand{\Linti}{L_{{\rm HX},i}}

\newcommand{\Lc}{L_{\rm{Corona}}}
\newcommand{\Lcn}{L_{\rm{Corona},44}}

\newcommand{\fmir}{f_{15\,\mu\rm{m}}}

\newcommand{\Lmir}{L_{15\,\mu\rm{m}}}
\newcommand{\Lmirn}{L_{15\,\mu\rm{m},43}}

\newcommand{\Lt}{L_{\rm{Torus}}}
\newcommand{\Ot}{\Omega_{\rm{Torus}}}

\newcommand{\Ld}{L_{\rm{Disk}}}
\newcommand{\Ldn}{L_{\rm{Disk},44}}
\newcommand{\Lbol}{L_{\rm{bol}}}
\newcommand{\Ledd}{L_{\rm{Edd}}}

\newcommand{\Fagn}{F_{\rm{AGN}}}

\newcommand{\Lir}{L_{12\,\mu\rm{m}}}
\newcommand{\Lx}{L_{\rm X}}

\newcommand{\Lsx}{L (2\,{\rm keV})}
\newcommand{\Lopt}{L (2500\AA)}

\newcommand{\integral}{{\em INTEGRAL}}
\newcommand{\spitzer}{{\em Spitzer}}

\shorttitle{Bolometric luminosity of Seyfert galaxies}
\shortauthors{Sazonov et al.}

\begin{document}

\author{S. Sazonov\altaffilmark{1,2}}

\author{S.P. Willner\altaffilmark{3}}

\author{A.D. Goulding\altaffilmark{3}}

\author{R.C. Hickox\altaffilmark{4}}

\author{V. Gorjian\altaffilmark{5}}

\author{M.W. Werner\altaffilmark{5}}

\author{E. Churazov\altaffilmark{2,1}}

\author{R. Krivonos\altaffilmark{2,1}}

\author{M. Revnivtsev\altaffilmark{1}}

\author{R. Sunyaev\altaffilmark{2,1}}

\author{C. Jones\altaffilmark{3}}

\author{S.S. Murray\altaffilmark{3,6}}

\author{A. Vikhlinin\altaffilmark{3,1}}

\author{A.C. Fabian\altaffilmark{7}}

\author{W.R. Forman\altaffilmark{2}}

\altaffiltext{1}{Space Research Institute, Russian Academy of
  Sciences, Profsoyuznaya 84/32, Moscow 117997, Russia}

\altaffiltext{2}{Max-Planck-Institut f\"{u}r Astrophysik,
  Karl-Schwarzschild-Str. 1, Garching 85741, Germany}

\altaffiltext{3}{Harvard-Smithsonian Center for Astrophysics, 60 Garden Street,
  Cambridge, MA 02138}

\altaffiltext{4}{Department of Physics and Astronomy, Dartmouth
  College, 6127 Wilder Laboratory, Hanover, NH 03755}

\altaffiltext{5}{MS 169-327, Jet Propulsion Laboratory, California Institute of
  Technology, 4800 Oak Grove Drive, Pasadena, CA 91109}

\altaffiltext{6}{Department of Physics and Astronomy, Johns Hopkins
  University, Baltimore, MD 21218}

\altaffiltext{7}{Institute of Astronomy, Madingley Road, Cambridge CB3
  0HA, United Kingdom}

\title{Contribution of the accretion disk, hot corona, and obscuring  
  torus to the luminosity of Seyfert galaxies: \integral\ and {\em
    SPITZER} observations}

\begin{abstract}
 We estimate the relative contributions of the supermassive black
  hole (SMBH) accretion disk, corona, and obscuring torus to the
  bolometric luminosity of Seyfert galaxies, using
  \spitzer\ mid-infrared (MIR) observations of a complete sample of 68
  nearby active galactic nuclei (AGNs) from the \integral\ all-sky
  hard X-ray (HX) survey. This is the first HX-selected (above 15~keV)
  sample of AGNs with complementary high angular resolution, high
  signal to noise, MIR data. Correcting for the host galaxy
  contribution, we find a correlation between HX and MIR luminosities:
  $\Lmir\propto\Lint^{0.74\pm0.06}$. Assuming that the observed MIR
  emission is radiation from an accretion disk reprocessed in a
  surrounding dusty torus that subtends a solid angle decreasing with
  increasing luminosity (as inferred from the declining
  fraction of obscured AGNs), the intrinsic disk luminosity, $\Ld$, is
  approximately proportional to the luminosity of the corona in the
  2--300~keV energy band, $\Lc$, with the $\Ld/\Lc$ ratio varying by a
  factor of 2.1 around a mean value of 1.6. This ratio is a factor of
  ${\sim} 2$ smaller than for typical quasars producing the cosmic X-ray
  background (CXB). Therefore, over three orders of magnitude
  in luminosity, HX radiation carries a large, and roughly comparable,
  fraction of the bolometric output of AGNs. We estimate the
  cumulative bolometric luminosity density of local AGNs at ${\sim}
  (1-3)\times 10^{40}$~erg~s$^{-1}$~Mpc$^{-3}$. Finally, the Compton
  temperature ranges between $kT_{\rm c}\approx 2$ and $\approx 6$~keV
  for nearby AGNs, compared to $kT_{\rm c}\approx 2$~keV for typical
  quasars, confirming that radiative heating of interstellar gas can
  play an important role in regulating SMBH growth.
\end{abstract}

 
%%%%%%%%%%%%%%%%%%%%%%
\section{Introduction}
\label{s:intro}
%%%%%%%%%%%%%%%%%%%%%%

Active galactic nuclei (AGNs) are extremely powerful sources of
electromagnetic radiation over many decades in frequency from
radiowaves to gamma-rays. According to the commonly accepted scenario,
an AGN shines due to accretion of gas onto a supermassive black hole
(SMBH) residing in a galactic nucleus. 

In Seyfert galaxies and quasars, most of the luminosity is emitted in 
the form of ultraviolet (UV) radiation generated in a geometrically
thin, optically thick accretion disk \citep{shasun73}, giving rise to
a ``big blue bump'' (BBB) in the spectral energy distribution (SED,
e.g., \citealt{malsar82}). Additional, higher energy radiation is
generated in a hot corona of the accretion disk (e.g.,
\citealt{haamar93}) and possibly also in collimated outflows (jets) of 
relativistic plasma, producing a hard X-ray (HX) peak in the SED. The
integrated (and redshifted) HX emission of all AGNs in the observable
Universe makes up the bulk of the cosmic X-ray background. There is
also a third, mid-infrared (MIR) peak in AGN SEDs (e.g.,
\citealt{barvainis87}), which arises from reprocessing of a
significant fraction of the disk's and some of the coronal radiation
in a torus of molecular gas and dust surrounding the inner accretion
flow. In fact, only in unobscured or ``type 1'' AGNs can all three
spectral components, the HX bump, the BBB, and the MIR bump, be
observed.  According to the unified model \citep{ant93}, these are
objects viewed through the funnel of the dusty torus. In contrast,
only the HX and MIR components are visible in the SEDs of obscured or
``type 2'' AGNs, because the torus is opaque to UV emission from the
accretion disk but transparent to coronal radiation at energies above
$\sim$15~keV (except in Compton-thick sources) and to its own infrared
emission (at least at wavelengths $\gtrsim20$~$\mu$m). All other
emission components, including broad- and narrow-line emission and
non-thermal radio and gamma-ray radiation are usually not significant
as regards their contribution to the angular-integrated bolometric
luminosity of AGNs; these components will therefore not be discussed
below.

To understand how electromagentic radiation is emitted and reprocessed
during accretion of matter onto SMBHs, it is crucial to explore i) in
what proportion the AGN luminosity is shared between the accretion
disk and its corona, ii) what fraction of the bolometric luminosity is
reprocessed in the torus, and iii) how these properties depend on
black hole mass and accretion rate. One also needs such information to
study the role of AGN feedback in regulating SMBH growth and galactic 
evolution. One of the proposed feedback mechanisms is photoionization 
and Compton heating of interstellar gas by AGN radiation (e.g.,
\citealt{cioost01,proetal08}), whose efficiency critically depends
on the AGN SED \citep{sazetal04,sazetal05}. Finally, information on
AGN SEDs can be used to derive bolometric corrections required to
reconstruct the cosmic history of SMBH accretion growth based on AGN
statistics provided by extragalactic surveys (e.g.,
\citealt{maretal04,merhei08}). 

Among all types of AGNs, the SEDs of unobscured high-luminosity
quasars have been studied most extensively (see, e.g.,
\citealt{elvetal94,ricetal06,shaetal11}). Their obscured counterparts
-- type~2 quasars -- have been explored to a much lesser degree,
although recent surveys have begun to find such objects in significant
numbers (e.g., \citealt{poletal06,hicetal07,lanetal09}). There is also
much uncertainty with respect to the SEDs of Seyfert galaxies, which
are typically less luminous than more distant quasars. 
The difficulty is that even in Seyfert 1s, the accretion disk 
emission is usually contaminated by host galaxy stellar emission in
visible bands and the BBB peaks in the observationally difficult
far-UV band (see, however, \citealt{scoetal04,vasfab07,vasfab09}).

The goal of the present study is to systematically assess the relative
contributions of the accretion disk, hot corona, and obscuring torus
to the bolometric luminosity of local Seyfert galaxies. To this end,
we i) cross-correlate the HX luminosities of nearby AGNs detected
during the all-sky survey of the {\em International Gamma-Ray
  Laboratory} (\integral, \citealt{winetal03}) with the MIR
luminosities of these objects measured by the {\em Spitzer Space
  Telescope} \citep{weretal04}, and ii) use the proportion of obscured
to unobscured AGNs to estimate the opening angle of dusty tori as a
function of luminosity. We then put our findings for nearby AGNs into
the broader context of cosmic SMBH growth by making a comparison with
distant quasars. 

Most previous relevant studies were based on AGN samples compiled
in a fairly arbitrary manner from optical and/or soft X-ray (below
10~keV) catalogs (e.g., \citealt{lutetal04,horetal06,honetal10}). In
these energy bands, AGNs can easily be missed due to absorption, as
powerful sources can become invisible when obscured by large amounts
of dust and cold gas in the torus and/or host galaxy. Furthermore, as 
already noted above, optical emission from relatively low-luminosity 
AGNs can be diluted against the background of a luminous galaxy (see
\citealt{mushotsky04} for a detailed discussion of AGN selection at
different wavelengths). 

The hard X-ray band, above $\sim 15$~keV, provides a census of AGNs  
that is far less biased with respect to the viewing orientation of 
the torus and is unbiased with respect to host galaxy
properties. There have been a few previous attempts
\citep{vasetal10,muletal11} of systematically studying the MIR
properties of HX selected AGNs using the {\em Swift} all-sky hard
X-ray survey \citep{tueetal10}. However, these studies either used
data from the IRAS all-sky photometric infrared survey, so that it was
impossible to reliably subtract the host galaxy contribution from the
AGN emission, or used high angular resolution \spitzer\ data but only
for statistically incomplete subsamples of {\em Swift} AGNs. Our
\integral\ sample is the first statistically complete, HX selected
sample of AGNs with complementary high angular resolution, high signal
to noise, MIR data. The extensive \spitzer\ coverage (3.6--38~$\mu$m)
available for the entire \integral\ sample makes this a unique data
set for studying SEDs of AGNs in the local Universe. 

%%%%%%%%%%%%%%%%%%%%%%%%%%%%%
\section{{\em INTEGRAL} AGN sample}
\label{s:sample}
%%%%%%%%%%%%%%%%%%%%%%%%%%%%%

Our study is based on the complete sample of AGNs
\citep{krietal07,sazetal07} detected in the 17--60~keV energy band by
the IBIS/ISGRI detector \citep{ubeetal03} aboard \integral\ during the
first three and a half years of the mission, from 2002 October until
2006 June. These observations compose a serendipitous all-sky HX
survey with the flux limit varying by a factor of a few over the
sky. We have excluded from the present analysis blazars (flat spectrum
radio quasars and BL Lac objects), a relatively rare subclass of AGNs
whose observed emission is believed to be dominated by a narrow,
strongly collimated component. We have also excluded AGNs located in
the ``zone of avoidance'' near the Galactic plane ($|b|<5^\circ$)
because there remain unidentified \integral\ sources in this region of
the sky while we wish our sample to be nearly 100\% complete to
minimize selection effects. 

The resulting set comprises 68 AGNs (Table~\ref{tab:agn}). 
In the first seven columns of Table~\ref{tab:agn} we have collected 
information on optical/radio AGN types, distances, HX fluxes and
luminosities, and X-ray absorption column densities ($\NH$). 
These data are mostly adopted from the original \integral\ catalog
\citep{krietal07,sazetal07} although some updates take into account
follow-up observations carried out since publication of the
catalog. In particular, thanks to recent X-ray observations by {\em
  Chandra}, {\em Swift}, and {\em XMM-Newton}, all of the 
previously missing $\NH$ values have now been estimated. 
All the reported absorption columns may be considered reliable   
because they are based on high signal-to-noise X-ray spectroscopic
data. We do not quote the uncertainties associated with the
$\NH$ values, because the information on absorption columns has
been compiled from various sources and in most cases the actual
uncertainty is likely dominated by systematic effects associated
with the particular spectral modeling procedure used.
In fact, multiple measurements taken for some AGNs at
different times and/or by different instruments sometimes yield $\NH$
values that differ from each other by more than their reported
uncertainties. We estimate that the total uncertainties associated
with $\NH$ columns for our sources are typically smaller than 30\%
and do not affect the present study in any significant way.

All but one of our AGNs are located
at low redshift ($z< 0.1$, the most distant one, IGR~J09446$-$2636,
being at $z=0.14$). 
For 18 nearby (closer than $\sim 40$~Mpc) Seyfert galaxies we have
adopted distance estimates from either \cite{tuletal09} or
\cite{tully88}; otherwise luminosity distances have been calculated
from the spectroscopic redshifts assuming a cosmology with
$\Omega_{\rm m}=0.3$, $\Omega_\Lambda=0.7$, and
$H_0=72$~km~s$^{-1}$~Mpc$^{-1}$.

\begin{deluxetable*}{lrcrrcrcrrrrr}
%\begin{deluxetable}{lrcrrcrcrrrrr}
\tabletypesize{\scriptsize}
\tablecaption{{\em INTEGRAL}--{\em Spitzer} AGN sample\label{tab:agn}} 
%\tablewidth{0pt}
\tablehead{
\colhead{Name} & \colhead{AGN\tablenotemark{a}} & \colhead{Ref} &
\colhead {$z$} & 
\colhead{$D$,} & \colhead{Ref} & \colhead{$\NH$,\tablenotemark{b}} & \colhead{Ref}  
& \multicolumn{2}{c}{17--60~keV} &
\multicolumn{3}{c}{15~$\mu$m} 
\\
\cline{9-10} \cline{11-13}
\colhead{} & \colhead{Class} & \colhead{} & \colhead{} &
\colhead{Mpc} & \colhead{} & \colhead{$10^{22}$~cm$^{-2}$} & \colhead{} &
\colhead{Flux, $10^{-11}$} & \colhead{$\log L$,} & \colhead{$f_\nu$,} &
\colhead{$\log\nu L_\nu$,} & \colhead{$\Fagn$}   
\\
\cline{2-3} \cline{5-6} \cline{7-8} 
\colhead{} & \colhead{} & \colhead{} & \colhead{} & \colhead{} &
\colhead{} & \colhead{} & \colhead{} & \colhead{erg~s$^{-1}$~cm$^{-2}$} &
\colhead{erg~s$^{-1}$} & \colhead{Jy} & \colhead{erg~s$^{-1}$} &
\colhead{} 
}
\startdata
\multicolumn{12}{c}{Clean sample (AGN dominated infrared sources)}\\
\hline
MRK 348          &   Sy2 &  & 0.0150 &  63.4 & & 30   &  &  $7.4\pm0.8$ & 43.55 & 0.413 & 43.60 & 0.97\\ 
MCG -01-05-047   &   Sy2 &  & 0.0172 &  72.8 & & 14   & 6&  $1.6\pm0.3$ & 43.02 & 0.107 & 43.14 & 0.73\\  
NGC 788          &   Sy2 &  & 0.0136 &  57.4 & & 40   &  &  $4.8\pm0.3$ & 43.28 & 0.216 & 43.23 & 1.00\\
LEDA 138501      &   Sy1 &  & 0.0492 & 213.3 & & $<1$ &  &  $4.0\pm0.7$ & 44.33 & 0.029 & 43.50 & 1.00\\ 
MRK 1040         & Sy1.5 &  & 0.0167 &  70.7 & & $<1$ &  &  $4.9\pm0.8$ & 43.46 & 0.574 & 43.84 & 1.00\\  
IGR J02343+3229  &   Sy2 &  & 0.0162 &  68.5 & & 2    & 7&  $3.9\pm0.6$ & 43.34 & 0.128 & 43.16 & 0.73\\ 
1H 0323+342      & NLSy1 &1 & 0.0610 & 266.7 & &$<1$  &  &  $2.7\pm0.5$ & 44.37 & 0.059 & 44.00 & 0.90\\
NGC 1365         & Sy1.8 &  & 0.0055 &  17.9 &5& 50   &  &  $3.3\pm0.7$ & 42.10 & 1.736 & 43.13 & 0.37\\
3C 111           & Sy1, BLRG &  & 0.0485 & 210.1 & & $<1$ &  &  $7.8\pm0.9$ & 44.62 & 0.137 & 44.16 & 1.00\\ 
ESO 033-G002     &   Sy2 &  & 0.0181 &  76.7 & &   1  &  &  $1.9\pm0.3$ & 43.14 & 0.300 & 43.63 & 1.00\\ 
IRAS 05078+1626  & Sy1.5 &  & 0.0179 &  75.8 & & $<1$ &  &  $5.9\pm0.8$ & 43.61 & 0.471 & 43.81 & 0.96\\
MRK 3            &   Sy2 &  & 0.0135 &  57.0 & &  100 &  &  $6.8\pm0.3$ & 43.43 & 1.352 & 44.02 & 1.00\\
MRK 6            & Sy1.5 &  & 0.0188 &  79.7 & &   5  &  &  $3.7\pm0.3$ & 43.45 & 0.361 & 43.74 & 0.96\\ 
ESO 209-G012     & Sy1.5 &  & 0.0405 & 174.4 & & $<1$ &  &  $1.7\pm0.2$ & 43.78 & 0.245 & 44.25 & 0.82\\
IRAS 09149$-$6206&   Sy1 &  & 0.0573 & 249.8 & & $<1$ &  &  $2.1\pm0.3$ & 44.19 & 0.645 & 44.98 & 1.00\\
MRK 110          & NLSy1 &  & 0.0353 & 151.5 & & $<1$ &  &  $5.9\pm1.1$ & 44.21 & 0.079 & 43.64 & 1.00\\
IGR J09446$-$2636& Sy1.5 &2 & 0.1425 & 658.2 & & $<1$ &  &  $3.9\pm0.7$ & 45.31 & 0.025 & 44.41 & 1.00\\ 
NGC 2992         &   Sy2 &  & 0.0077 &  30.5 &4&  1   &  &  $5.1\pm0.4$ & 42.75 & 0.627 & 43.15 & 0.48\\
MCG -5-23-16     &   Sy2 &  & 0.0085 &  35.7 & &  2   &  &  $9.8\pm0.8$ & 43.17 & 1.111 & 43.53 & 1.00\\
NGC 3081         &   Sy2 &  & 0.0080 &  32.5 &4& 50   &  &  $4.6\pm0.6$ & 42.77 & 0.490 & 43.09 & 0.96\\
ESO 263-G013     &   Sy2 &  & 0.0333 & 142.7 & & 40   &  &  $2.2\pm0.4$ & 43.72 & 0.086 & 43.62 & 1.00\\ 
NGC 3227         & Sy1.5 &  & 0.0039 &  20.6 &4& $<1$ &  &  $9.1\pm0.8$ & 42.66 & 0.697 & 42.85 & 0.63\\
NGC 3281         &   Sy2 &  & 0.0107 &  45.1 & & 150  &  &  $3.9\pm0.6$ & 42.98 & 1.327 & 43.81 & 1.00\\
IGR J10386$-$4947& Sy1.5 &  & 0.0600 & 262.1 & &   1  &  &  $1.5\pm0.2$ & 44.08 & 0.081 & 44.12 & 1.00\\
IGR J10404$-$4625&   Sy2 &  & 0.0239 & 101.7 & &   3  &  &  $2.1\pm0.3$ & 43.42 & 0.171 & 43.63 & 0.91\\
NGC 3783         &   Sy1 &  & 0.0097 &  38.5 &4& $<1$ &  & $12.3\pm1.9$ & 43.34 & 1.037 & 43.57 & 1.00\\ 
IGR J12026$-$5349&   Sy2 &  & 0.0280 & 119.5 & &   2  &  &  $2.4\pm0.3$ & 43.62 & 0.439 & 44.18 & 0.92\\
NGC 4151         & Sy1.5 &  & 0.0033 &  20.3 &4&    8 &  & $47.4\pm0.4$ & 43.37 & 2.804 & 43.44 & 1.00\\
MRK 50           &   Sy1 &  & 0.0234 &  99.5 & & $<1$ &  &  $1.3\pm0.2$ & 43.19 & 0.026 & 42.80 & 1.00\\
NGC 4388         &   Sy2 &  & 0.0084 &  16.8 &4& 40   &  & $17.9\pm0.3$ & 42.78 & 0.776 & 42.72 & 0.93\\
{\bf NGC 4395}         & Sy1.8 &  & 0.0011 &   4.6 &5&  2   &  &
$1.6\pm0.3$  & 40.59 & 0.013 & 39.83 & 0.86\\
NGC 4507         &   Sy2 &  & 0.0118 &  49.7 & &  60  &  & $10.9\pm0.5$ & 43.51 & 0.859 & 43.71 & 0.95\\
NGC 4593         &   Sy1 &  & 0.0090 &  39.5 &4& $<1$ &  &  $5.9\pm0.3$ & 43.04 & 0.477 & 43.25 & 0.78\\
ESO 323-G077     & Sy1.2 &  & 0.0150 &  63.4 & &  30  &  &  $2.8\pm0.3$ & 43.13 & 0.689 & 43.82 & 0.78\\
IGR J13091+1137  & XBONG&  & 0.0251 & 106.9 & &  90  &  &  $3.5\pm0.4$ & 43.68 & 0.092 & 43.40 & 1.00\\
IGR J13149+4422  &   Sy2 &  & 0.0366 & 157.2 & &5     & 7&  $2.2\pm0.4$ & 43.81 & 0.207 & 44.09 & 0.98\\ 
Cen A            & Sy2, NLRG &  & 0.0018 &   3.6 &5& 11   &  & $56.0\pm0.3$ & 41.94 &2.184  & 41.83  & 0.94\\
MCG -6-30-15     & Sy1.2 &  & 0.0077 &  32.4 & &$<1$  &  &  $3.6\pm0.4$ & 42.66 & 0.483 & 43.09 & 1.00\\
MRK 268          &   Sy2 &  & 0.0399 & 171.8 & &30    & 8&  $1.7\pm0.3$ & 43.79 & 0.127 & 43.95 & 0.84\\
IC 4329A         & Sy1.2 &  & 0.0160 &  67.7 & &$<1$  &  & $16.1\pm0.5$ & 43.95 & 1.517 & 44.22 & 1.00\\ 
NGC 5506         & Sy1.9 &  & 0.0062 &  28.7 &4&  3   &  & $13.3\pm0.7$ & 43.12 & 1.739 & 43.54 & 1.00\\
IGR J14552$-$5133& NLSy1 &  & 0.0160 &  67.7 & &$<1$  &  &  $1.4\pm0.2$ & 42.89 & 0.173 & 43.28 & 0.95\\
IC 4518A         &   Sy2 &  & 0.0157 &  66.4 & &  10  &  &  $2.4\pm0.2$  & 43.11 & 0.410 & 43.64 & 0.87\\
WKK 6092         &   Sy1 &  & 0.0156 &  65.9 & &$<1$  &  &  $1.7\pm0.2$ & 42.94 & 0.068 & 42.85 & 1.00\\
IGR J16185$-$5928&   Sy1 &  & 0.0350 & 150.1 & &$<1$  & 9&  $1.7\pm0.2$ & 43.67 & 0.035 & 43.28 & 1.00\\
ESO 137-G034     &   Sy2 &  & 0.0092 &  38.7 & &$\gtrsim 100$&10& $1.7\pm0.2$ & 42.48 & 0.231 & 42.92 & 0.95\\
IGR J16482$-$3036&   Sy1 &  & 0.0313 & 133.9 & &$<1$  &  &  $2.6\pm0.2$ & 43.75 & 0.039 & 43.22 & 1.00\\ 
NGC 6221         &   Sy2 &  & 0.0050 &  19.4 &4& 1    &  &  $1.9\pm0.3$ & 41.93 & 0.875 & 42.90 & 0.46\\
IGR J16558$-$5203& Sy1.2 &  & 0.0540 & 234.9 & &$<1$  &  &  $2.9\pm0.2$ & 44.29 & 0.212 & 44.45 & 0.91\\  
NGC 6300         &   Sy2 &  & 0.0037 &  14.3 &4& 25   &  &  $4.7\pm0.4$ & 42.06 & 0.891 & 42.64 & 0.94\\
IGR J17418$-$1212&   Sy1 &  & 0.0372 & 159.8 & & $<1$ &  &  $2.6\pm0.3$ & 43.89 & 0.173 & 44.03 & 0.94\\
3C 390.3         & Sy1, BLRG &  & 0.0561 & 244.4 & & $<1$ &  &  $6.2\pm0.6$ & 44.64 & 0.147 & 44.32 & 1.00\\
IGR J18559+1535  &   Sy1 &  & 0.0838 & 372.3 & & $<1$ &  &  $2.3\pm0.2$ & 44.58 & 0.096 & 44.50 & 1.00\tablenotemark{c}\\
1H 1934$-$063    & NLSy1 &3 & 0.0106 &  44.6 & & $<1$ &  &  $1.8\pm0.3$ & 42.63 & 0.514 & 43.39 & 0.94\\ 
NGC 6814         & Sy1.5 &  & 0.0052 &  22.8 &4& $<1$ &  &  $4.7\pm0.4$ & 42.47 & 0.178 & 42.35 & 1.00\\
Cyg A            & Sy2, NLRG &  & 0.0561 & 244.4 & & 20   &  &  $5.8\pm0.3$ & 44.62 & 0.323 & 44.67 & 1.00\\
MRK 509          & Sy1.2 &  & 0.0344 & 147.5 & & $<1$ &  &  $5.5\pm0.8$ & 44.16 & 0.395 & 44.31 & 0.89\\
NGC 7172         &   Sy2 &  & 0.0087 &  33.9 &4&   13 &  &  $6.0\pm0.5$ & 42.92 & 0.349 & 42.98 & 0.65\\
MR 2251$-$178    &   Sy1 &  & 0.0640 & 280.4 & &$<1$  &  &  $4.8\pm0.5$ & 44.65 & 0.119 & 44.35 & 1.00\\
NGC 7469         & Sy1.2 &  & 0.0163 &  68.9 & &$<1$  &  &  $4.7\pm0.8$ & 43.43 & 1.552 & 44.25 & 0.54\\
MRK 926          & Sy1.5 &  & 0.0469 & 203.0 & &$<1$  &  &  $3.6\pm0.5$ & 44.25 & 0.139 & 44.14 & 1.00\\ 
\hline
\multicolumn{12}{c}{Starburst dominated infrared sources}\\
\hline
NGC 1142         &   Sy2 &  & 0.0288 & 123.0 & & 45   &  &  $4.6\pm0.4$ & 43.92 & 0.065& 43.37 & $\lesssim 0.5$\\
ESO 005-G004     &   Sy2 &  & 0.0062 &  22.4 &4&100   &11&  $2.5\pm0.5$ & 42.18 & 0.196 & 42.37 & $\lesssim 0.5$\\
IGR J07563$-$4137&   Sy2 &  & 0.0210 &  89.1 & &$<1$  &  &  $1.2\pm0.2$ & 43.07 & 0.046 & 42.95 & $\lesssim 0.5$\\
NGC 4945         &   Sy2 &  & 0.0019 &   3.8 &5& 200  &  & $19.9\pm0.4$ & 41.54 & 1.872 & 41.81 & $\lesssim 0.5$\\
IGR J14561$-$3738&   Sy2 &  & 0.0246 & 104.7 & &$\gtrsim 100$&12& $1.4\pm0.3$ & 43.27 & 0.051 & 43.13 & $\lesssim 0.5$\\
MCG +04-48-002   &   Sy2 &  & 0.0142 &  60.0 & & 50   &  &  $3.3\pm0.6$
& 43.16 & 0.268 & 43.37 & $\lesssim 0.5$ \\
\hline
\multicolumn{12}{c}{Compton-thick objects}\\
\hline
NGC 1068         &   Sy2 &  & 0.0038 &14.4    &4&$\gtrsim1000$& &
$1.9\pm0.3$ & 41.67 & 15\tablenotemark{d} & 43.87 & 1.00
\enddata

\tablerefs{(1) object also exhibits some blazar properties
  \citep{zhoetal07}, (2) \cite{masetal08}, (3) \cite{rodetal00}, (4)
  \cite{tully88}, (5) \cite{tuletal09}, (6) \cite{lanetal07}, (7)
  \cite{rodetal08}, (8) {\em XMM-Newton} data, (9) \cite{maletal08}, (10)
  \cite{maletal09}, (11) \cite{uedetal07}, (12) \cite{sazetal08a}.
}
\tablenotetext{a}{AGN optical/radio classes are from \cite{sazetal07} unless a 
reference is given: Sy1--Sy2 -- Seyfert galaxy, NLSy1 --
narrow-line Seyfert 1 galaxy, BLRG -- broad-line radio galaxy, 
NLRG --  narrow-line radio galaxy, XBONG -- X-ray bright optically 
normal galaxy. The LLAGN NGC~4395 is marked in bold.}
\tablenotetext{b}{X-ray absorption columns are from
    \cite{sazetal07} unless a reference is given.} 
\tablenotetext{c}{Poor IRS SL ($\lambda<14$~$\mu$m) data due to
  inaccurate slit position, spectral shape at $\lambda>14$~$\mu$m indicates
  negligible starburst contribution.}
\tablenotetext{d}{The source is saturated in low-resolution IRS data, 
  the flux density is from \cite{masetal06}, AGN fraction is assumed
  to be 100\%.}

\end{deluxetable*} 
%\end{deluxetable} 

The \integral\ AGN sample has HX (17--60~keV) luminosities ranging
over almost five orders of magnitude, from $4\times 10^{40}$
(NGC~4395) to $2\times 10^{45}$ (IGR~J09446$-$2636) erg~s$^{-1}$.
Therefore, this is a representative, hard X-ray selected sample of
nearby AGNs, mostly Seyfert galaxies, although our $\sim 10$ most
luminous objects may be better referred to as nearby quasars, because
their HX luminosities exceed $10^{44}$~erg~s$^{-1}$.

One special object in the sample is the Seyfert 1.8 galaxy NGC~4395, a
famous low-luminosity AGN (LLAGN) sometimes referred to as a ``dwarf
Seyfert nucleus''. It appears to be a quite atypical Seyfert galaxy in
terms of its black hole mass, luminosity, and variability properties (see, e.g.,
\citealt{moretal99,petetal05,vauetal05}). It is therefore possible
that the properties of its dusty torus (if there is any) are also different
from typical Seyfert galaxies and quasars. We thus treat this object
separately from the rest of the sample in performing the HX--MIR
cross-correlation analysis (\S\ref{s:hx_mir}).

As noted above, the HX selection (17--60 keV) implies that there is
almost no bias from absorption. Our AGN sample is not sensitive to 
photoabsorption (and we have thus not corrected the
measured HX fluxes for line-of-sight absorption)
as long as the column density of the gas is less than
a few $10^{24}~{\rm cm^{-2}}$ or equivalently the Thomson optical
depth is less than a few; at even larger column densities, the flux
from a source drops considerably at all X-ray energies. 
The Seyfert 2 galaxy NGC~1068 distinguishes itself from the rest of 
the sample because it is the only significantly Compton-thick AGN
(having $\NH\gtrsim 10^{25}$~cm$^{-2}$, \citealt{matetal00}). We
therefore exclude NGC~1068 from our baseline HX--MIR cross-correlation
analysis but discuss its properties in comparison with Compton-thin
sources (\S\ref{s:hx_mir}, \S\ref{s:torus_disk_corona}).

%%%%%%%%%%%%%%%%%%%%%%%%%%%%%%%%%%%%%%%%%%%%%%%%%%
\section{{\em SPITZER} observations and data reduction}
\label{s:analysis}
%%%%%%%%%%%%%%%%%%%%%%%%%%%%%%%%%%%%%%%%%%%%%%%%%%

More than half of the \integral\ sample consists of well-known Seyfert
galaxies, many of which have been targets of observational campaigns
with \spitzer. For the remaining part, largely represented by AGNs
discovered by \integral, we carried out short \spitzer\ observations
(Program ID 50763) consisting of 3.6--8~$\mu$m imaging with IRAC and
low-resolution MIR spectroscopy with IRS; in addition, far-infrared
photometry was performed with MIPS for a subset of objects. Our
proprietary and publicly available archival data together provide
complete coverage of the \integral\ sample with \spitzer\ at
3.6--38~$\mu$m.  

All of the sources in our hard X-ray selected
  sample would have been 
  robustly detected by \spitzer\ even if they had been 1--3 orders of
  magnitude fainter. Hence, our sample is not limited by MIR flux and
  any correlations derived between the HX and MIR luminosities can be
  considered representative of the local AGN population without
  significant bias.

%%%%%%%%%%%%%%%%%
\subsection{IRS}
\label{s:irs}

We used the InfraRed Spectrograph (IRS, \citealt{houetal04}) on
\spitzer\ to obtain low-resolution spectra of our objects. Our program's
observations, for a total of 30 AGNs, were done in spectral mapping
mode using the Short-Low (SL) and Long-Low (LL) IRS modules. Each of
these modules has first- and second-order sub-slits (SL1, SL2, LL1,
and LL2) with widths of 3.7, 3.6, 10.7, and 10.5~arcsec,
respectively. We used SL1, LL1, and LL2 for the entire sample and SL2 for a
subset of objects. The resulting spectra thus cover a range from
either 5.2 or 7.5~$\mu$m up to 38~$\mu$m. Observations with SL1 and SL2
consisted of one cycle of 6 pointings with a ramp duration of 6~s
with two 19$\arcsec$ steps in the slit direction and three
1.8$\arcsec$ steps in the dispersion direction. For LL1 and LL2, one
cycle of 3 pointings with a ramp duration of 6~s and a step size of
42$\arcsec$ in the slit direction was implemented.

For those sources that were not covered by our \spitzer\ program, we
used archival low-resolution IRS data: mapping mode observations
for 14 AGNs and staring mode observations for another 23 AGNs. Almost
all of the archival mapping mode observations have the following
setup: one cycle of 13 pointings with a step size of 1.8$\arcsec$ in
the dispersion direction for SL1 and SL2 and 5 pointings with a step
size of 4.85$\arcsec$ in the dispersion direction for LL1 and LL2, the
ramp duration being 6~s. For the archival staring mode observations,
the ramp times and numbers of cycles vary from one object to another.

In the analysis of mapping mode observations, we used basic calibrated data
(BCD), extracted from the \spitzer\ Science Center pipeline (versions 
S18.0.1 and newer for our program's observations and versions S15.3.0
and newer for the archival observations). For each order of a given
IRS module and for each position of a given source within the slit, we first 
produced a background image. For our program's observations, this was done 
by averaging over 2D spectra obtained in significantly ($\sim 19\arcsec$ 
for SL and $\sim 42\arcsec$ for LL) off-source positions, whereas for the 
archival observations, a similar averaging was done over 2D spectra obtained 
in the other order of a given IRS module. The background image was then 
subtracted from the on-source 2D spectrum. Then, a 1D spectum of the source 
was obtained using the \spitzer\ IRS Custom Extraction software (SPICE)
by applying the regular extraction algorithm, which uses an aperture that
gradually increases with wavelength in accordance with the telescope's
point spread function (e.g., the aperture width is 7.2$\arcsec$ at 6~$\mu$m
and 36.6$\arcsec$ at 27~$\mu$m). Finally, an averaging over a set of 
1D spectra extracted in different source positions within the slit was done.
The staring mode spectra were obtained simply by averaging over nod-subtracted 
post-BCD spectra. These were derived with the same regular extraction algorithm 
as was used in our analysis of mapping mode observations. 

Although the data reduction procedure described above is fairly simplistic and
does not fully exploit the potential of mapping mode observations (which are
available for nearly two thirds of our sample), it is adequate for the 
purposes of our HX--MIR cross-correlation study, as confirmed by a comparison
with an alternative, more detailed analysis of IRS data for a subset
of \integral\ sources (see \S\ref{s:fmir_uncert}).

The Compton-thick Seyfert 2 galaxy NGC~1068 is an extremely bright
($\sim 15$~Jy at 15~$\mu$m) infrared source, which caused saturation
of IRS. We therefore quote in Table~\ref{tab:agn} its MIR flux
estimate based on a compilation of high angular resolution, infrared
observations \citep{masetal06}.

%%%%%%%%%%%%%%%%%
\subsection{IRAC}
\label{s:irac}

We used images obtained by the InfraRed Array Camera (IRAC,
\citealt{fazetal04}) to determine source flux densities at 3.6, 4.5,
5.8, and 8.0~$\mu$m. This wavelength range partially overlaps with
that covered by IRS spectroscopy, enabling direct comparison between
the spectroscopic and photometric results and providing an extension
of spectra to shorter wavelengths.

Our IRAC program included 35 \integral\ sources. Observations were done
in high dynamic range (HDR) mode. Specifically, a combination of a
0.6~s frame and a 12~s frame was repeated in 9 random dithers for each
source. For the rest of the sample, we made use of archival
observations, for which the number and duration of frames varied from
source to source. Since most of our objects are bright infrared
sources, our analysis was in most cases based on stacking the 0.6~s
frames. For the 5 weakest sources ($\lesssim 10$~mJy at
3.6--8.0~$\mu$m), including LEDA~138501, IGR~J09446$-$2636,
ESO~263-G013, IGR~J10386$-$4947, and NGC~4395, to improve the
accuracy, we determined the fluxes by stacking the long (2, 12, or
30~s) frames. This was also done for 3C~111 and Mrk~3 despite their
relative brightness because the observations were not done in HDR
mode. We verified that none of the sources was saturated.

We analyzed post-BCD data using the standard point source extraction 
package APEX, part of the MOPEX software. Post-BCD images are adequate 
for this work because their main deficiency, poor artifact correction, 
is unimportant for 0.6-s frames. We estimated source fluxes by 
integrating the surface brightness in different apertures with radii
between 2.4$\arcsec$ to 12$\arcsec$ and correcting for flux leakage
outside the aperture under the assumption of a point-like
source. Although this procedure is inaccurate for measuring fluxes of
spatially extended sources, it is good enough to indicate the presence
of extended host galaxy emission as a significant difference between
fluxes measured in large and small apertures (see
Appendix~\ref{s:app_irs_irac}).  
 
The existing IRAC photometric measurements for the Compton-thick
Seyfert 2 galaxy NGC~1068 proved to be saturated and hence were not used.

%%%%%%%%%%%%%%%%%%%%%%%%%%%%%%%%%%%%%%%%%%%%%%%%%%%%%%%%
\section{Infrared spectra: AGN MIR emission}
\label{s:spectra}
%%%%%%%%%%%%%%%%%%%%%%%%%%%%%%%%%%%%%%%%%%%%%%%%%%%%%%%%

\begin{figure*}
\centering
\includegraphics[bb=20 140 585 700,width=\textwidth]{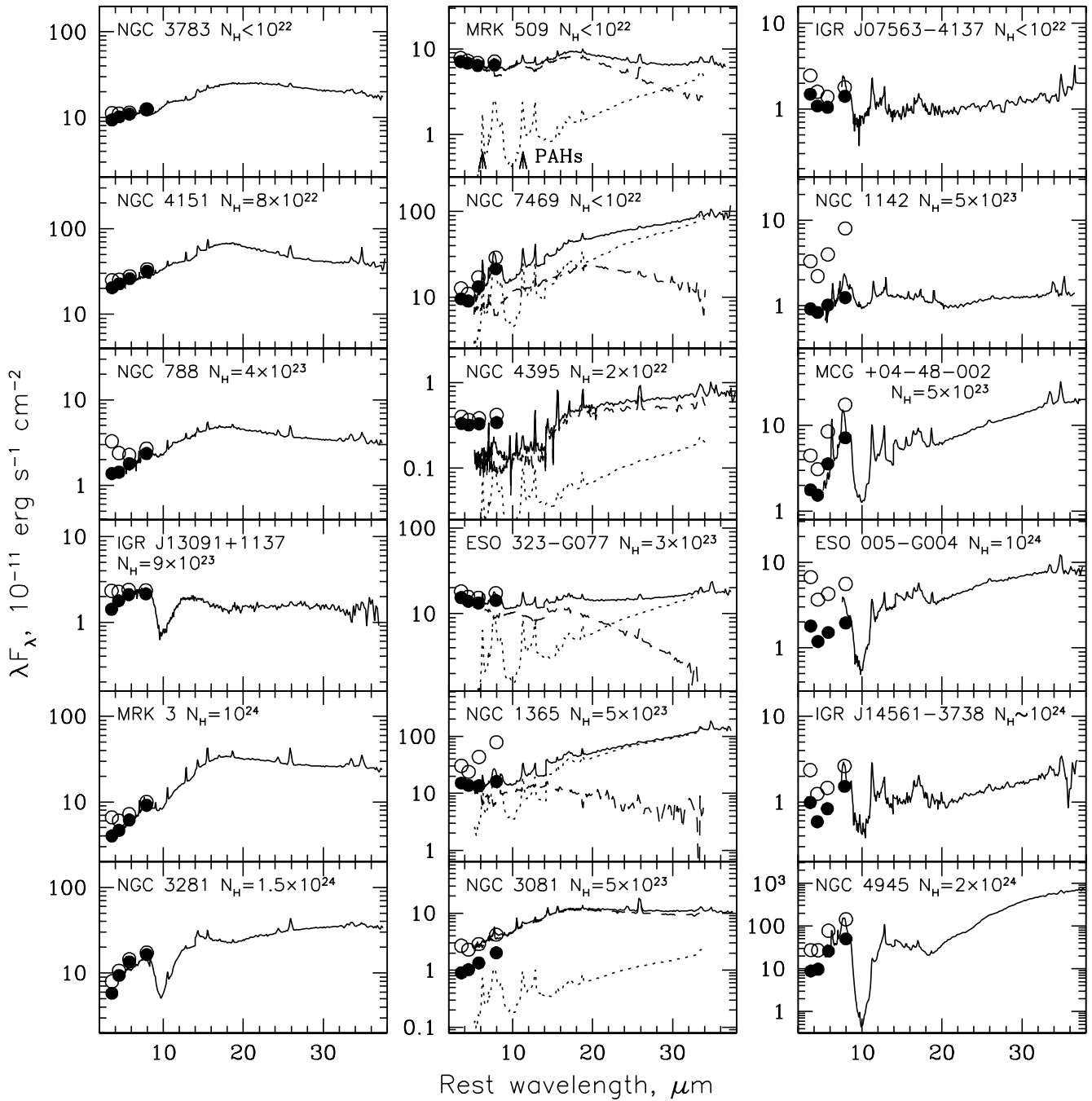}
\caption{Examples of \spitzer\ spectra of \integral\ AGNs. In each panel,
  the solid line shows the low-resolution IRS spectrum, and the filled 
  and open circles represent fluxes at 3.6, 4.5, 5.8, and 8.0~$\mu$m
  derived from IRAC images in 2.4$\arcsec$ and 12$\arcsec$ apertures,
  respectively (corrected for flux leakage outside the aperture
  assuming a point source). Also, the X-ray absorption columns are
  indicated. {\sl Left column:} Spectra that are clearly dominated by
  dust emission associated with the active nucleus, with a weak or
  absent starburst contribution. {\sl Middle column:} Spectra showing
  a noticeable contribution of MIR emission from dust associated with 
  star formation. The dotted line shows the starburst
  component, estimated by fitting the starburst template to the 6.2
  and 11.3~$\mu$m PAH lines (indicated in the upper panel). The dashed
  line shows the AGN contribution, found as the difference
  between the total spectrum and the starburst component; it may still
  be contaminated at short wavelengths by stellar and accretion disk
  emission. {\sl Right column:} Starburst dominated spectra, for which
  extraction of an AGN component is practically impossible. Note the
  much broader flux range covered by the spectrum of NGC~4945 compared
  to the other sources.
}
\label{fig:spectra}
\end{figure*}

Infrared emission from dust associated with star formation in the host
galaxy can provide a significant contribution to AGN MIR spectra. Therefore,
to study torus emission, we need to estimate and subtract the star
formation contribution from the \spitzer\ data. We observe clear
signatures of star formation in many of our IRS spectra. These
include polycyclic aromatic hydrocarbon (PAH) emission features and the
$\nu F_\nu$ continuum rising toward the far-infrared. We therefore
modelled the measured spectra by a sum of starburst and AGN
components, similarly to a number of previous studies (e.g.,
\citealt{netetal07,muletal11}).

Figure~\ref{fig:spectra} shows examples of IRS spectra with negligible,
significant, and strong star formation contribution as deduced using
the fitting procedure described below. At short wavelengths, we also
show IRAC photometric fluxes measured in 2.4$\arcsec$ and
12$\arcsec$ apertures.

%%%%%%%%%%%%%%%%%%%%%%%%%%%%%%%%%%%%%%%%%%%%%%%%%%%%%%%%%%%%
\subsection{Spectral decomposition using a starburst template}
\label{s:template}

We adopted the starburst template from \cite{braetal06}\footnote{
http://www.strw.leidenuniv.nl/$\sim$brandl/SB\_template.html}, which
is an average over low-resolution IRS spectra of a dozen nearby
($D\lesssim 100$~Mpc) starburst galaxies. This template is
well suited for our analysis because it was obtained by low-resolution
IRS spectroscopy, similarly to the spectra studied here. 

We normalized starburst components in our objects based on the
observed strength of PAH lines, which are believed to be a generic
signature of star formation \citep{rocetal91}. The presence of an AGN
in a star forming galaxy may lead to a weakening of the PAH spectral
features because PAH molecules can be destroyed by hard AGN radiation
\citep{voit92}. However, the importance of this effect is
controversial (see, e.g., \citealt{smietal07b,odoetal09,saletal10}),
and we have assumed that the shape of the starburst spectral component is not
affected by the presence of a central AGN.

Our analysis consisted of the following steps. First, we fitted the spectra
around (typically within $\pm 0.6$~$\mu$m of) the 6.2~$\mu$m and
11.3~$\mu$m PAH lines by a sum of a linear continuum and a
Gaussian. We then compared the derived PAH line fluxes with the
corresponding values for the starburst template, which yielded two
independent estimates of the amplitude of the star-formation component. 
The average of these two values was then adopted as the normalization
of the starburst template. On average, the coefficients implied by the
6.2~$\mu$m and 11.3~$\mu$m features proved to be consistent with each 
other, although there is $\sim 40$\% scatter around the 1:1 ratio of
the two coefficients. This indicates that there are $\sim 20$\%
systematic uncertainties in the derived amplitudes of starburst
components for our objects (this issue is further discussed in
\S\ref{s:fmir_uncert} below). If there were no IRS data for the 
6.2~$\mu$m feature (i.e., only first-order SL data at $\gtrsim 7.5$~$\mu$m 
were available), we used the flux of the 11.3~$\mu$m line to normalize the 
starburst component. If the observed PAH features proved to be strong enough
(depending on the source brightness, we required the PAH equivalent
widths, EW, to be larger than 0.01--0.02~$\mu$m, as compared to EW=0.45
and 0.55~$\mu$m for the 6.2~$\mu$m and 11.3~$\mu$m bands, respectively, in the
starburst template), we subtracted the estimated starburst
contribution from the total spectrum to derive the AGN
component. Otherwise, we considered star formation contamination
insignificant and did not perform any subtraction.

We applied an additional correction to 13 spectra that contained detectable
PAH features and exhibited a significant ($>$10--20\%, depending on
the source brightness) discontinuity near 14~$\mu$m, where  
the short-wavelength segment measured in the 3.7$\arcsec$-wide SL1
slit connects to the long-wavelength segment measured in the
10.5$\arcsec$-wide LL2 slit. Such a ``jump'' is mostly likely caused
by an extended source, i.e., it cannot be due to the AGN. In these
objects, the long-wavelength ($\gtrsim 14$~$\mu$m) part of the
starburst component was rescaled to make the AGN component smooth
across the SL--LL boundary. We did not make such a correction for
NGC~4395, by far the weakest infrared source in our sample (with an
estimated flux density of 13~mJy at 15~$\mu$m), despite the  
apparent presence of a significant SL--LL discontinuity in its
spectrum, because of its low statistical quality (see
Fig.~\ref{fig:spectra}). Furthermore, our comparison with available 
high-resolution spectroscopy for this object (see \S\ref{s:components}
below) indicates that the SL slit was not positioned sufficiently
accurately on the nucleus of NGC~4395, which might have caused an
artificial discontinuity at 14~$\mu$m in the low-resolution IRS spectrum.

In principle, we could also use another known strong PAH feature, at
7.7~$\mu$m, for estimating the contribution of star formation. However, 
this band overlaps with the high-ionization [NeVI] 7.65~$\mu$m line,
which can be bright in AGNs \citep{stuetal02} and is impossible to
separate from the PAH feature in low-resolution IRS spectra. Moreover,
the 7.7~$\mu$m feature might be significantly affected by AGN radiation
\citep{smietal07b,odoetal09}.

%%%%%%%%%%%%%%%%%%%%%%%%%%%%%%%%%%%%%%%%%%%%%%%%%%%%%%%%%%%%%%%%%%%%%%%%
\subsection{AGN dominated (clean sample) vs. starburst dominated sources}
\label{s:components}

In agreement with previous studies (e.g.,
\citealt{weeetal05,bucetal06,shietal06,deoetal09,wuetal09}), we
observe a large variety of infrared spectral shapes among Seyfert  
galaxies. However, those spectra dominated by dust reprocessed emission 
generated by black hole accretion (see examples in the left
column of Fig.~\ref{fig:spectra}), rather than by star formation,
almost invariably peak (when plotted in $\nu F_\nu$ units) at
$\sim$15--20~$\mu$m, in good agreement with models of dusty tori
heated by a central source of UV radiation 
(e.g., \citealt{dulvan05,honetal06,nenetal08,aloetal11}). Furthermore,
the AGN components of those IRS spectra with inferred significant  
starburst contamination (see the middle column of Fig.~\ref{fig:spectra}) 
prove to be similar to the spectra of ``pure'' AGNs. In particular,
most of the former also peak at 15--20~$\mu$m.  However, since our
procedure of estimating the star formation contribution based on the
strength of PAH features becomes progressively less reliable with
increasing wavelength, there is much uncertainty in the deduced AGN
spectral contributions at $\lambda\gtrsim 20$~$\mu$m. All these
findings are similar to the results of previous attempts to
decompose \spitzer\ spectra of quasars and Seyfert galaxies into AGN
and starburst components (e.g., \citealt{netetal07,muletal11}).

The IRS spectra of six objects, NGC~1142, ESO~005-G004, IGR~J07563$-$4137,
NGC~4945, IGR~J14561$-$3738, and MCG+04-48-002 (see the right column of panels
in Fig.~\ref{fig:spectra}) closely resemble the starburst template. We
found it practically impossible to distinguish AGN and host galaxy
components in these starburst dominated sources and therefore excluded
them (see Table~\ref{tab:agn}) from most of the subsequent
analysis. Interestingly, most of our starburst 
dominated objects are strongly X-ray absorbed AGNs ($\NH\sim
10^{24}$~cm$^{-2}$). One may speculate that i) large supplies of cold
gas and dust associated with starburst activity in a galactic nucleus
facilitate the formation of a dense central obscuring torus, and/or ii)
part of the X-ray absorption is caused by cold gas tracing star
formation in the galaxy and located outside a parsec-scale AGN torus.

The remaining 61 objects (with NGC~1068 excluded for 
being a Compton-thick source) compose a ``clean'' sample for our
subsequent analysis. As concerns the LLAGN NGC~4395, since its (low
signal to noise) low-resolution IRS spectrum leaves doubts as to the
presence of a significant starburst contribution (we estimate it at
$\sim 14$\% at 15~$\mu$m), we have also analyzed available
high-resolution IRS data, following the methods described aby
\cite{gouale09}. While the derivation of an accurate continuum shape,
using only high-resolution IRS, is complicated by the tip-tilt effects
of the individual echelle orders, the high-resolution
MIR spectrum of NGC~4395 is characterized almost entirely by an
AGN-produced broken power law with little or no evidence (${\rm EW}\ll 
0.1$~$\mu$m) for superposed PAH features. Hence, NGC~4395
is clearly AGN dominated and thus should be part of our clean
sample. Nevertheless, as noted before, we still distinguish this
``dwarf Seyfert'' from the rest of the sample during our HX--MIR
cross-correlation analysis because it might represent a physically
different class of AGNs.

%%%%%%%%%%%%%%%%%%%%%%%%%%%%%%%%%%%%%%%%%%
\subsection{15~$\mu$m flux and luminosity}
\label{s:flux15}

We have just seen (Fig.~\ref{fig:spectra} and
\S\ref{s:components}) that after subtraction of the star formation
contribution, AGN MIR continua have an approximately constant
shape. This suggests that it should be possible to estimate bolometric
luminosities of AGN obscuring tori using their flux
densities measured at a single MIR wavelength. We have chosen to use for this
purpose the rest-frame $\lambda=15$~$\mu$m. Specifically,
$f_\nu$~(15~$\mu$m) was determined by averaging a given spectrum over
the wavelength range 14.7--15.2~$\mu$m. There are several reasons
behind this choice. First, $\lambda=15$~$\mu$m is approximately where
AGN torus emission peaks. Second, since IRS SL2 data are not available
for some of our sources, we can only use wavelengths $\lambda\gtrsim
8$~$\mu$m for the whole sample. Third, wavelengths $\lambda\gtrsim
20$~$\mu$m are disfavored because cool dust emission associated with
star formation becomes more important with increasing wavelength and frequently
dominates far-infrared and even mid-infared spectra of
Seyferts. Finally, there are no strong emission lines or absorption
features within $\sim 0.5$~$\mu$m on either side of 15~$\mu$m.

The last three columns of Table~\ref{tab:agn} present the total
measured flux densities ($f_\nu$) and corresponding luminosities ($\nu L_\nu$)
at 15~$\mu$m as well as estimated fractions of AGN emission in the total 
flux at 15~$\mu$m, $\Fagn$. Statistical uncertainties for the infrared
fluxes and luminosities are negligibly small. For the six starburst
dominated objects, we assume that $\Fagn\lesssim 50$\%. In the
subsequent analysis, the starburst subtracted MIR flux of AGNs is defined as
\beq
\fmir=\Fagn\nu f_\nu~(15~\mu{\rm m}).
\label{eq:fmir}
\eeq
The corresponding AGN luminosity is defined as
\beq
\Lmir=\Fagn\nu L_\nu~(15~\mu{\rm m}).
\label{eq:lmir}
\eeq

%%%%%%%%%%%%%%%%%%%%%%%%%%%%%%%%%%%%%%%%%%%%%%%%%%
\subsection{Uncertainty in AGN MIR flux estimates}
\label{s:fmir_uncert}

The main potential source of systematic uncertainty in our estimation
of AGN fluxes at 15~$\mu$m is the use of a fixed starburst spectral
template from \cite{braetal06}. In reality, the spectral properties of
infrared emission from dust heated by starbursts may vary from one
galaxy to another, and some authors (e.g., \citealt{muletal11}) have
attempted to take this diversity into account in separating
AGN and host galaxy components for Seyfert galaxies. 

A crude estimate of the systematic uncertainty associated with our use
of a fixed spectral template was already made in \S\ref{s:template}
using the difference in the normalizations of starburst spectral
components determined using the 6.2 and 11.3~$\mu$m PAH
features. Namely, starburst component amplitudes 
could be estimated by our fitting procedure to within $\sim 20$\%.
This implies that for AGN dominated sources, i.e., those objects with 
$\Fagn\gtrsim 50$\%, the AGN fluxes at 15~$\mu$m are also estimated to within 
$\sim 20$\%, i.e., to better than 0.1~dex in log space. 

To better understand the uncertainties associated with our estimates
of $\fmir$, we performed an alternative spectral analysis based on a set of
starburst spectral templates for a subsample of our
objects. Specifically, we selected a representative subset (16 
objects) of AGN-dominated, mixed, and starburst-dominated
sources. This comparison sample is equally divided between
sources with IRS-staring and mapping data. In \S\ref{s:template}, we
assumed that the SL---LL discontinuity observed in the MIR spectra of
some of these AGNs arises due to extended host galaxy emission which
provides an additional contribution to the larger aperture LL
spectrum. However, it has also been suggested in previous studies that
this discontinuity between the spectral orders derives from an IRS
detector effect; for the purposes of comparison, we imposed this
assumption for our sub-sample. Furthermore, for those sources with IRS
mapping data, we combined the rogue-pixel cleaned BCD images and
extracted nuclear spectra using the 3D spectral reduction program CUBISM
\citep{smietal07a}, which is used widely in recent MIR AGN literature
(e.g., \citealt{daletal09,gouale09,diarie10,petetal11,aloetal12}). We
used DecompIR \citep{muletal10} to deconvolve the MIR spectra for our 
comparison sample using a set of empirical and theoretical starburst
templates (e.g., \citealt{gouetal11}) and assumed an absorbed broken
power law to model the AGN component. The derived AGN fractions and
fluxes prove to be entirely consistent with the values established
by our baseline analysis (\S\ref{s:irs}, \S\ref{s:template}) for
those sources with $\Fagn>0.5$ (the average scatter in the derived
$\fmir$ is $\pm8$\%). However, for those sources with 
$\Fagn<0.5$ (i.e., the spectra appear starburst dominated), the
spectral fits become strongly dependent on the imposed starburst
templates, and the scatter in the measured AGN flux increases to a
factor of $\sim 2$. 

As an additional check of our IRS spectral measurements, we can use the
IRAC imaging data available for all of our AGNs. 
For the vast majority of the sources, the absolute values of flux
densities measured by IRAC and IRS in the overlapping spectral region
below 8~$\mu$m are in good mutual agreement (see
Fig.~\ref{fig:spectra}), especially when the smallest (2.4$\arcsec$)
IRAC aperture is used (recall that the IRS SL slits 
have similar widths, 3.6--3.7$\arcsec$). However, there are a few
sources for which there is a significant discrepancy between the
spectroscopic and photometric fluxes at $\lambda\lesssim
8$~$\mu$m. Since the corresponding IRAC and IRS observations were
separated by several years in time, these flux differences probably
indicate significant intrinsic variability of AGN torus emission,
especially at shorter wavelengths
\citep{sugetal06,kisetal09,trietal09,kozetal10}. Also, as shown in 
Appendix~\ref{s:app_irs_irac}, the detection of significant extended
emission at 8~$\mu$m by IRAC in many sources is fully consistent with
our conclusions about the host-galaxy contamination of IRS spectra.  

Finally, we compared our results for 16 AGNs with higher angular resolution
observations from \cite{ganetal09}. We find excellent agreement (see
Fig.~\ref{fig:gandhi} and discussion in \S\ref{s:previous})  
between the fluxes derived from the very different observational data
sets, which suggests that host galaxy contamination of our measured
values of AGN MIR fluxes is minimal. 

We conclude that the combined statistical and systematic uncertainty
in our AGN 15~$\mu$m-flux estimates for the clean sample (i.e., AGN
dominated sources) is probably less than 0.1~dex. This
uncertainty proves to be small in comparison with the intrinsic
scatter in the HX--MIR flux and luminosity correlations (see
\S\ref{s:hx_mir} below). We further discuss the potential influence of starburst
contamination on our derived correlations in \S\ref{s:tests}.

%%%%%%%%%%%%%%%%%%%%%%%%%%%%%%%%%%%%%%
\section{HX--MIR luminosity relation}
\label{s:hx_mir}
%%%%%%%%%%%%%%%%%%%%%%%%%%%%%%%%%%%%%%

Before comparing the luminosities of AGN structural
components (accretion disk, corona, and obscuring torus), which will
be the subject of the next section (\S\ref{s:torus_disk_corona}), we
first perform a cross-correlation analysis of AGN luminosities
measured in the HX and MIR bands by \integral\ and \spitzer, 
respectively. The results of this analysis may also be interesting in
its own right for any studies addressing links between X-ray and
infrared emission in AGNs.

\begin{figure}
\centering
\includegraphics[bb=15 150 550 680, width=\columnwidth]{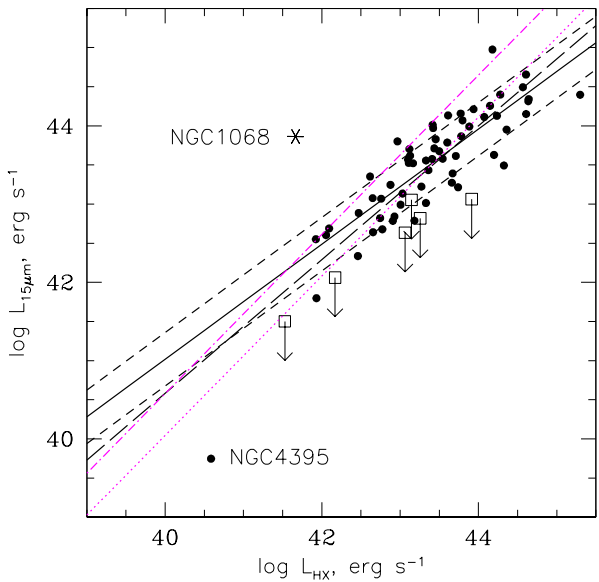}
\caption{Luminosity scatter plot of $\Lmir$ vs. $\Lint$. Filled
  circles represent AGNs from the clean sample (excluding six
  starburst dominated AGN). The black solid line shows the best-ftting
  power law ($\Lmir$ as a function of $\Lint$) for these objects
  excluding NGC~4395 (eq.~[\ref{eq:xray_15m_lum}]), while the two
  black dashed lines show this dependence multiplied and divided by
  2.19, the rms scatter around the mean trend. The black long-dashed
  line shows the best-fitting relation for the total clean sample
  including NGC~4395. The correlation parameters are listed in
  Table~\ref{tab:fits}. Empty squares denote the 6 starburst dominated
  sources, for which an AGN fraction $\Fagn<50$\% at 15~$\mu$m is
  assumed. Also shown is the Compton-thick Seyfert 2 galaxy
  NGC~1068. The magenta dotted line shows the result of fitting
  $\Lint$ as a function of $\Lmir$ by a power law,
  eq.~(\ref{eq:15m_xray_lum}), for the clean sample excluding
  NGC~4395. The magenta dash-dotted line shows the same dependence
  corrected for the Malmquist bias, eq.~(\ref{eq:15m_xray_lum_malm}). 
  } 
\label{fig:xray_15m_lum}
\end{figure}

Figure~\ref{fig:xray_15m_lum} shows the scatter plot of $\Lmir$
vs. $\Lint$, where $\Lint$ is the luminosity in the 17--60~keV energy
band and $\Lmir$ was defined in eq.~(\ref{eq:lmir}). In computing
luminosities from fluxes, we neglected uncertainties associated with
source distances.

\begin{deluxetable*}{lllllllll}
%\tabletypesize{\scriptsize}
\tablecaption{Results of HX--MIR cross-correlation analysis
 for the clean sample of AGNs and its subsamples: $\Lmirn=a\Lintn^b$
\label{tab:fits}
} 

\tablehead{
\colhead{Sample} & \colhead{Number} & \colhead{$a$} &  \colhead {$b$} &
\colhead{rms,} & \multicolumn{2}{c}{Spearman} &
\multicolumn{2}{c}{Pearson}
\\
\cline{6-9}
\\
\colhead{} & \colhead{of objects} & \colhead{} & \colhead{} &
\colhead{dex} & \colhead{$\rho$} & \colhead{$P_{\rm null}$} & \colhead{$r$} & \colhead{$P_{\rm null}$} 
}
\startdata
{\bf Without NGC 4395} & {\bf 60} & $\bf 1.7\pm0.2$ & $\bf
0.74\pm0.06$ & {\bf 0.34} & {\bf 0.85} & $\bf 7\times 10^{-18}$  &
{\bf 0.85} & $\bf 8\times 10^{-18}$ \\   
All & 61 & $1.4\pm0.2$ & $0.85\pm0.06$ & 0.39 & 0.86 &
$9\times 10^{-19}$ & 0.88 & $3\times 10^{-20}$\\  
$\Fagn\ge0.9$ & 46 & $1.7\pm0.3$ & $0.72\pm0.07$ & 0.35 & 0.82 &
$5\times 10^{-12}$ & 0.84 & $5\times 10^{-13}$ \\  
Without NGC 4395, total MIR fluxes & 60 & $2.0\pm0.2$ &
$0.68\pm0.06$ & 0.35 & 0.83 & $1.2\times 10^{-16}$ & 0.82 & $8\times 10^{-16}$\\ 
$z<0.02$ (without NGC 4395) & 35 & $1.8\pm0.2$ & $0.93\pm0.10$ & 0.29
& 0.84 & $3\times 10^{-10}$ & 0.85 & $1.3\times 10^{-10}$\\ 
$z>0.02$ & 25 & $1.7\pm 0.8$ & $0.69\pm0.16$ & 0.38 & 0.67 &
$2\times 10^{-4}$ & 0.66 & $3\times 10^{-4}$\\ 
Sy1s and NLSy1s & 33 & $2.1\pm0.4$ & $0.69\pm0.09$ & 0.37 & 0.81 &
$1.2\times 10^{-8}$ & 0.80 & $2\times 10^{-8}$\\
Sy2s (without NGC 4395) & 27 & $1.7\pm0.2$ & $0.85\pm0.09$ & 0.30 &
0.85 & $3\times 10^{-8}$ & 0.88 & $1.7\times 10^{-9}$
\enddata

\end{deluxetable*}

Considering the clean AGN sample without NGC~4395 and fitting
$\Lmir$ as a function of $\Lint$ (computing the linear regression in
log-log space), we find a strong, non-linear
correlation between MIR and HX luminosities (see Table~\ref{tab:fits}):
\beq
\Lmirn=(1.7\pm0.2) \Lintn^{0.74\pm0.06},
\label{eq:xray_15m_lum}
\eeq
where the luminosities are measured in units of
$10^{43}$~erg~s$^{-1}$. The rms scatter of $\Lmir$ values around the
mean trend is 0.34~dex.  

As can be seen in Fig.~\ref{fig:xray_15m_lum}, NGC~4395 is a clear
outlier from the luminosity correlation, with its MIR luminosity being almost 2
orders of magnitude below the $\Lint$--$\Lmir$ trend described by
eq.~(\ref{eq:xray_15m_lum}). If we consider this LLAGN together with
the rest of the clean sample, the slope of the correlation increases
from 0.74 to 0.85 (see Table~\ref{tab:fits}), although formally the
change is not significant.

Fig.~\ref{fig:xray_15m_lum} also shows the six starburst dominated
sources, assuming that their AGN fractions $\Fagn<50$\%. Surprisingly, 
the upper limits to the 15~$\mu$m fluxes of AGN components for all these 
objects lie below the best-fitting relation for AGN dominated sources
(eq.~[\ref{eq:xray_15m_lum}]). At least in some cases, this
behavior is likely caused by attenuation of the intrinsic MIR emission
from the nucleus in the obscuring torus and in the surrounding galaxy (see
  \citealt{gouetal12}). In 
particular, the IRS spectrum of NGC~4945 (Fig.~\ref{fig:spectra})
exhibits a very deep silicate absorption trough at 10~$\mu$m, which,
assuming the standard composition of interstellar dust
\citep{draine03} and the simplest scenario of an infrared source
surrounded by a shell of dust, suggests that the neighboring continuum
emission at $\sim 15$~$\mu$m should be attenuated by a factor of $\sim
3$--5. In reality, depending on the actual distribution of dust in the
nucleus and body of the galaxy, the AGN MIR emission can be absorbed
even more strongly than suggested by the depth of the 10~$\mu$m
trough.

Finally, Fig.~\ref{fig:xray_15m_lum} shows the Compton-thick 
Seyfert 2 galaxy NGC~1068, which is a clear outlier from the
correlation between $\Lint$ and $\Lmir$ described by
eq.~(\ref{eq:xray_15m_lum}). This result is expected because intrinsic
hard X-ray emission is strongly absorbed in this object, and it is only 
the infrared signal that reveals the true power of this AGN. In fact,
the discrepancy between the general trend and the position of NGC~1068
on the $\Lint$--$\Lmir$ diagram suggests that its true X-ray
luminosity is two orders of magnitude higher than measured by
\integral, i.e., $L_{\rm HX,~unabsorbed}\sim
10^{44}$~erg~s$^{-1}$. This estimate is consistent with values
reported in the literature (e.g., \citealt{matetal00}).

\begin{figure}
\centering
\includegraphics[bb=15 150 550 680, width=\columnwidth]{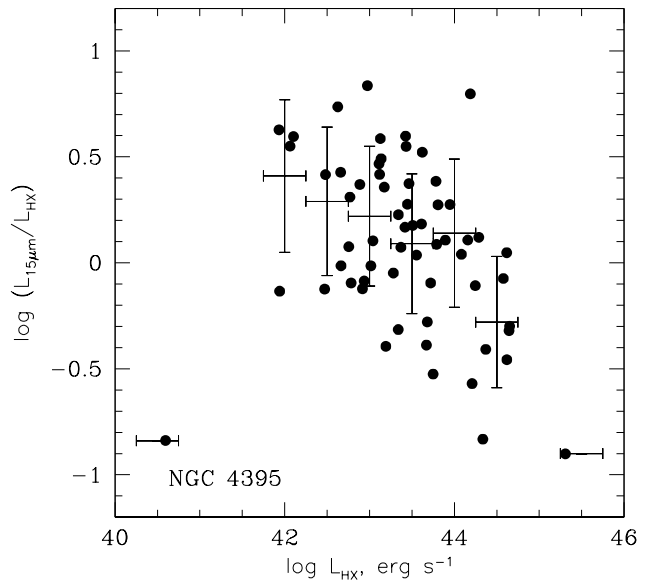}
\caption{$\Lmir/\Lint$ ratio as a function of $\Lint$ for the
  clean sample of AGNs. The uncertainties associated with the
  luminosities are less than $\sim 0.1$~dex. Averages over
  0.5~dex-wide bins in $\Lint$ are also shown, with the vertical error
  bars illustrating the rms scatter of individual measurements within bins. 
}
\label{fig:xray_15m_lum_ratio}
\end{figure}

The mean trend described by eq.~(\ref{eq:xray_15m_lum}) suggests that
the MIR/HX luminosity ratio decreases with increasing
$\Lint$. This can be better seen in Fig.~\ref{fig:xray_15m_lum_ratio},
which shows the $\Lmir/\Lint$ ratio as a function of $\Lint$. Grouping
our clean sample into 0.5~dex-wide bins in $\Lint$ shows that the
$\Lmir/\Lint$ ratio decreases from $\sim 1$--3 at $\Lint\sim
10^{42}$--$10^{43}$~erg~s$^{-1}$ to $\sim 0.3$--1 at $\Lint\sim
10^{44}$--$10^{45}$~erg~s$^{-1}$, although the ``dwarf Seyfert''
NGC~4395 is a clear outlier from this trend.

%%%%%%%%%%%%%%%%%%%%%%%%%%%%%%%%%%%%%%%%%%
\subsection{Robustness of the correlation}
\label{s:tests}

The derived $\Lint$--$\Lmir$ relation, eq.~(\ref{eq:xray_15m_lum}), 
makes it possible to predict the HX luminosity for a given MIR
luminosity. 
The combination of three facts, i) that our AGN sample is hard X-ray
selected, ii) that this sample is not limited by sensitivity in the
MIR band, and iii) that the correlation has been derived by fitting
$\Lmir$ as a function of $\Lint$, ensures that this relation
reproduces the intrinsic correlation between $\Lint$ and $\Lmir$ for
the local AGN population without any bias. 

To further test the robustness of the derived trend, we repeated our
$\Lint$--$\Lmir$ cross-correlation analysis for various subsamples of
AGNs. First, one may ask whether our procedure of separating AGN and 
starburst spectral components significantly affects the results. 
To address this issue, we computed the correlation for 46 strongly
AGN dominated sources -- those objects in which the AGN component 
accounts for at least 90\% of the total emission at 15~$\mu$m 
(i.e., $\Fagn>0.9$). The result (see Table~\ref{tab:fits}) is very close 
to the correlation found for the total clean sample from which
NGC~4395 is excluded (eq.~[\ref{eq:xray_15m_lum}]). In addition, we repeated the
analysis for the clean sample (without NGC~4395) using total measured
15~$\mu$m fluxes instead of AGN fluxes (i.e., setting $\Fagn=1$). The
amplitude of the correlation increased by $\sim 20$\%, 
obviously due to the unsubtracted contribution of starburst emission,
but  the slope changed by less than $1\sigma$ from 0.74 to 0.68. These tests
demonstrate that the correlation between $\Lint$ and $\Lmir$ is not
significantly affected by the details of our spectral analysis of IRS data. 

We next repeated the analysis separately for nearby 
($z<0.02$, 35 objects, excluding NGC~4395) and distant ($z>0.02$,
25 objects) sources from the clean sample. The correlation, in 
particular the slope of $0.69\pm0.16$, derived for
the distant subsample (see Table~\ref{tab:fits}) is fully
consistent with the correlation found for the total sample 
(eq.~[\ref{eq:xray_15m_lum}]). Since the $z>0.02$ set, owing to the \integral\ 
detection limit, is represented by luminous AGNs only, with 
$\Lint\sim 10^{43}$--$10^{45}$~erg~s$^{-1}$, this result also  
implies that the slope of the high-luminosity part of the HR--MIR 
correlation is not significantly different from the trend found 
over a broader range of luminosities. However, the slope, $0.93\pm 0.10$, 
determined for the nearby ($z<0.02$) subsample, mainly consisting
of lower luminosity AGNs with $\Lint\sim 10^{42}$--$10^{44}$~erg~s$^{-1}$,
is somewhat different from the general trend, but this difference is less 
than $2\sigma$ significant. 

\begin{figure}
\centering
\includegraphics[bb=15 150 550 680, width=\columnwidth]{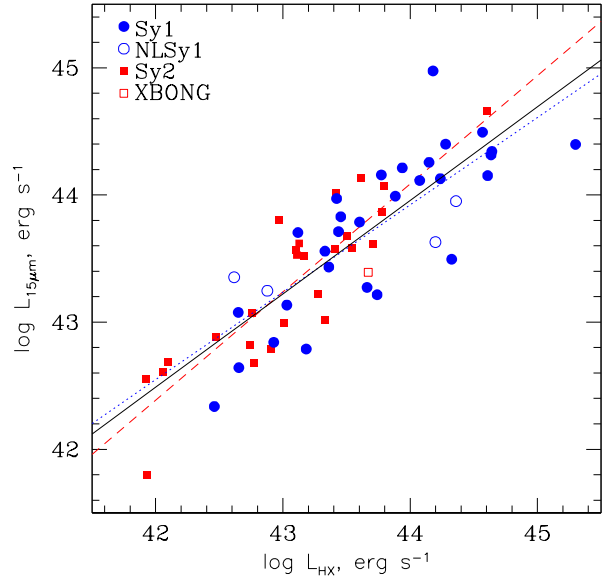}
\caption{Scatter plot of $\Lmir$ vs. $\Lint$ for the clean sample of
  AGNs excluding NGC~4395. Blue filled circles and red squares denote
  Seyfert 1s and Seyfert 2s, respectively. Also indicated are
  narrow-line Seyfert 1s (empty circles) and the X-ray bright
  optically normal galaxy IGR~J13091+1137 (empty square). The solid black,
  dotted blue, and dashed red lines show the best-fitting power laws
  for the clean sample (without NGC~4395), Sy1s, and Sy2s, respectively
  (Table~\ref{tab:fits}).
}
\label{fig:xray_15m_lum_types}
\end{figure}

Finally, we repeated our analysis for different types of AGNs, namely
Seyfert 1s (Sy1, including the intermediate types 1.2 and 1.5) and Seyfert
2s (Sy2, including the intermediate types 1.8 and 1.9). The derived
relations (Table~\ref{tab:fits}) are consistent with each other and with
eq.~(\ref{eq:xray_15m_lum}). As can be seen from
Fig.~\ref{fig:xray_15m_lum_types}, Sy1s and Sy2s do not distinguish 
themselves on the $\Lint$--$\Lmir$ diagram, nor do narrow-line Seyfert 1
galaxies occupy a distinct region of this diagram. Finally, there is
no significant dependence of the $\Lmir/\Lint$ ratio on the X-ray
absorption column density except for the clear separation of the
extremely Compton-thick source NGC~1068 from the rest of the sample.

We conclude that the correlation between HX and MIR luminosities
described by eq.~(\ref{eq:xray_15m_lum}) is robust, although 
there is a weak indication that the slope of the correlation is not
constant and decreases with increasing AGN luminosity. 

For some applications, one may also be interested in knowing the
distribution of $\Lint$ for a given $\Lmir$, rather than $\Lmir$ as a
function of $\Lint$. We have therefore also computed
(Appendix~\ref{s:app_inverse}) the inverse linear regression, i.e.,
$\log\Lint$ as a function of $\log\Lmir$, for our clean sample
excluding NGC~4395. As can be seen in Fig.~\ref{fig:xray_15m_lum},
this relation is different from the dependence of $\Lmir$ on $\Lint$.

%%%%%%%%%%%%%%%%%%%%%%%%%%%%%%%%%%%%%%%%%%
\subsection{Comparison with previous work}
\label{s:previous}

\cite{muletal11} have studied infrared properties of nearby AGNs detected
in the 14--195~keV energy band by {\em Swift}/BAT. This sample,
although not statistically complete, is similar to our \integral\ sample
in that it is hard X-ray selected. Using a subsample of AGNs having
both \spitzer/IRS spectroscopic and IRAS photometric data,
\cite{muletal11} developed and tested a procedure, based on a set of
starburst spectral templates, that make it possible to separate AGN
and starburst contributions to the infrared flux using IRAS four-band
photometry only. They then applied this procedure to a sample of 44
BAT AGNs and found that $L_{12\,\mu\rm{m},43}=(2.4\pm 0.4) L_{\rm
  14-195~keV,43}^{0.74\pm 0.13}$. This result is in excellent
agreement with our eq.~(\ref{eq:xray_15m_lum}). 

On the other hand, \cite{ganetal09} have reported a near
proportionality between 2--10~keV ($\Lx$) and 12~$\mu$m 
luminosities for Seyfert galaxies using high angular resolution
infrared observations: their best estimate is
$\Lir\propto\Lx^{1.11\pm0.07}$. This result seems to contradict our
conclusion that the $\Lmir/\Lint$ ratio decreases with increasing
luminosity.

A number of factors might contribute to this discrepancy, but the
most important one appears to be the difference in sample
luminosities. A difference in galaxy weighting makes at most a minor
difference. In their preferred regression procedure,
\cite{ganetal09} took into account individual uncertainties in X-ray
and infrared luminosities. However, the X-ray uncertainties were
estimated in a rather arbitrary way, taking into account long-term
variability for some sources but not for others. This led to
significantly different weights ascribed to different sources in
fitting. In our view, for the problem at hand, it is preferable to use
a standard linear regression procedure in log--log space giving equal
weights to all the sources in a sample. In fact, \cite{ganetal09} did
perform such an analysis and obtained a somewhat flatter dependence
$\Lir\propto\Lx^{1.02\pm0.07}$, which is, however, still significantly
steeper than the $\Lmir\propto\Lint^{0.74\pm0.06}$ relation found
here.

Another potentially important factor is the use of 2--10~keV X-ray
luminosities by \citet{ganetal09} vs. our use of HX luminosities. Due
to our hard X-ray selection and the resulting insensitivity to absorption
effects, we are able to determine HX luminosities directly from
17--60~keV fluxes measured by \integral. In comparison, the
\cite{ganetal09} sample contains a large number of significantly
absorbed ($\NH\gtrsim 10^{23}$~cm$^{-2}$) sources whose intrinsic
2--10~keV luminosities were estimated through model-dependent analysis
of X-ray spectra or, in some cases, even using [OIII] optical line
fluxes. We have compared the HX and X-ray luminosities for 16 Seyfert
galaxies (excluding the Compton thick Seyfert NGC~1068) which are
present in both the \integral\ and \citet{ganetal09} samples. The
data, spanning the $\Lint$ range from $9\times 10^{41}$~erg~s$^{-1}$
(Cen~A) to $1.4\times 10^{44}$~erg~s$^{-1}$ (Mrk~509), prove to be
consistent with $\Lint$ being proportional to $\Lx$. The small scatter 
(0.23~dex) associated with this correlation can be fully attributed to
the uncertainties in the $\Lx$ values as estimated by
\citet{ganetal09}, whereas the mean ratio $\Lint/\Lx\approx 1.5$ may
be considered typical for AGNs and is only slightly larger than the
ratio (1.2) corresponding to a fiducial AGN spectrum used in our
analysis below (\S\ref{s:corrections}). Furtheremore, systematic
studies based on AGNs from the \integral\ \citep{deretal12} and {\em
  Swift}/BAT \citep{winetal09} hard X-ray surveys have not revealed a 
significant dependence of the HX/X-ray flux ratio on luminosity. We
thus conclude that the use of $\Lx$ by \citet{ganetal09} vs. our use
of $\Lint$ is unlikely to lead to a significant difference between the
results of the corresponding X-ray--infrared cross-correlation
analyses. 

\begin{figure}
\centering
\includegraphics[bb=15 150 550 680, width=\columnwidth]{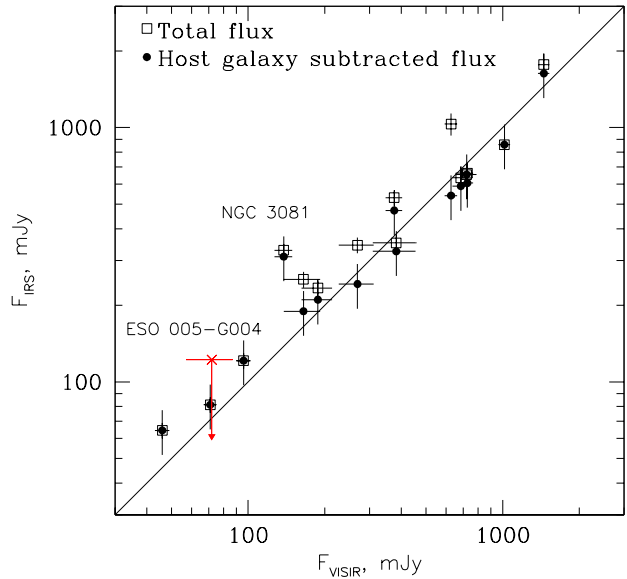}
\caption{Comparison of fluxes measured at $\lambda\approx
    12$~$\mu$m by VLT/VISIR \citep{ganetal09} and \spitzer/IRS. The
    black empty squares are total IRS fluxes, and the black dots are
    starburst-subtracted fluxes, to which systematic uncertainties of
    20\% are ascribed. The red cross with the arrow is the total flux
    for ESO~005-G004, for which IRS cannot confidently resolve the AGN
    component. The 1:1 dependence is shown with the solid line.}
\label{fig:gandhi}
\end{figure}

\begin{figure}
\centering
\includegraphics[bb=15 150 550 650, width=\columnwidth]{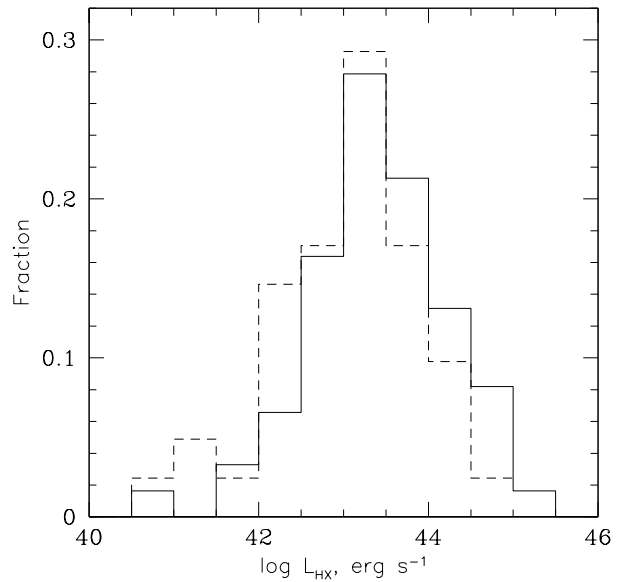}
\caption{Distribution of AGNs from our clean sample (solid line) and the
  \citet{ganetal09} one (dashed line) over the HX luminosity. For the
  latter sample we assumed that $\Lint=1.5\Lx$.} 
\label{fig:gandhi_lhx}
\end{figure}

Differences in the infrared data analysis might also play a role. Our
study is based on \spitzer/IRS spectroscopy and uses a template based
separation of AGN and starburst spectral components. \citet{ganetal09}
use narrow-filter photometry near rest-frame 12~$\mu$m with the
VISIR instrument on VLT, which provides a significantly better angular
resolution compared to \spitzer/IRS and hence presumably minimizes
host galaxy contamination. We have again used the overlapping sample
of 16~Seyferts to check if there are any systematic differences associated
with these two approaches. Specifically, we used our IRS spectra and
derived fluxes within the VISIR filters that were used by 
\citet{ganetal09}, which differ from object to object. As shown in
Fig.~\ref{fig:gandhi}, although the total fluxes measured by IRS
spectra are somewhat higher than those measured by VISIR, our standard
correction for the host galaxy contamination brings both data sets to  
nearly perfect agreement. A significant difference (a factor of $\sim
2$) is only observed for NGC~3081, but it seems natural to expect some
deviations in a sample of 16 objects, given that
\spitzer/IRS and VLT/VISIR observations were not
simultaneous (for the same reason, the 8~$\mu$m fluxes measured by
IRAC and IRS differ significantly for some of our AGNs, see
\S\ref{s:fmir_uncert}). Furthermore, there is no trend   
with either distance or luminosity, although the comparison
sample spans distances from 3.6~Mpc (Cen~A) to 174~Mpc
(ESO~209-G012). We conclude that the differences in the infrared data
analysis do not significantly bias our results with respect to
those of \citet{ganetal09}. 

Perhaps most importantly, \cite{ganetal09} used a heterogeneous
sample of 42 Seyfert galaxies, whereas we use a statistically complete
and somewhat larger sample. As a result, the \citet{ganetal09} sample
is significantly shifted to lower luminosities relative to ours 
(Fig.~\ref{fig:gandhi_lhx}): e.g., the corresponding fractions of AGNs 
with $\Lint>10^{44}$~erg~s$^{-1}$ are 12\% and 38\%. As was
dicusssed in \S\ref{s:tests}, dividing our sample into two subsets,
$z<0.02$ and $z>0.02$, represented by relatively low and high
luminosity AGNs, respectively ($\sim 10^{42}$--$10^{44}$ vs. $\sim
10^{43}$--$10^{45}$~erg~s$^{-1}$), tentatively suggests that the slope
of the X-ray--infrared correlation changes from $0.93\pm 0.10$ to
$0.69\pm 0.16$ as the AGN luminosity increases. The \citet{ganetal09}
sample effectively probes the luminosity range $\Lint\lesssim
10^{44}$~erg~s$^{-1}$, similar to our $z<0.02$ subsample, and
the slopes inferred for these two data sets are in
satisfactory agreement with one another. This suggests that the results
of \citet{ganetal09} and the present work are actually consistent with 
each other. 

We conclude that further studies using larger, well-defined samples of
AGNs are required to clarify if the slope of the X-ray--infrared
correlation depends on luminosity, as tentatively
suggested by the existing data.

%%%%%%%%%%%%%%%%%%%%%%%%%%%%%%%%%%%
\section{Torus vs. disk and corona}
\label{s:torus_disk_corona}
%%%%%%%%%%%%%%%%%%%%%%%%%%%%%%%%%%%

The unified model posits that a torus of molecular gas
and dust subtending a solid angle $\Ot$ (if viewed from the SMBH)
intercepts optical, UV, and soft X-ray radiation from the central
accretion flow and converts it into thermal infrared
emission. Therefore, assuming that the central source of radiation is 
isotropic, the total luminosity of the torus is expected to be 
\beq
\Lt\approx\frac{\Ot}{4\pi}\Ld,
\label{eq:ltor}
\eeq
where $\Ld$ is the luminosity of the accretion disk, presumably emitted between
$\lambda\sim 1$~$\mu$m (NIR) and $E\sim 2$~keV (soft X-rays). These boundaries
usually separate the MIR, BBB, and HX components (see \S\ref{s:intro})
in the SEDs of type~1 AGNs (see, e.g.,
\citealt{elvetal94,sazetal04}). Physically, the 1~$\mu$m boundary 
marks the onset of thermal emission from hot dust at sublimation
temperature ($\sim 1500$~K), whereas accretion disk emission is
expected to peak in the near- or far-UV bands in quasars and Seyfert galaxies  
(e.g., \citealt{shasun73,hubetal01}). Therefore, the chosen energy boundaries 
for $\Ld$ ensure that virtually all of the accretion disk luminosity
is accounted for.

Depending on the column density through the torus, it can also
reprocess a fraction $\lesssim\Ot/(4\pi)$ of the higher energy
(2--10~keV) luminosity emitted by a hot corona of the accretion
disk. We have neglected this contribution in eq.~(\ref{eq:ltor}),
first because the torus's optical depth may be significantly smaller
than unity for 5--10~keV X-rays, in contrast to the softer emission
from the accretion disk, and also because we expect the X-ray (below
10~keV) luminosity of the corona to be much lower than the bolometric
luminosity of the accretion disk. This last assumption will be
verified below upon completion of our cross-correlation
analysis. Finally, it is assumed that none of the hard X-ray emission
(above 10~keV) is absorbed within the AGN, which is a reasonable
assumption expect for very Compton-thick objects such as NGC~1068. 

As demonstrated below, \spitzer\ and \integral\ data together make it
possible to estimate the luminosity ($\Lt$) and the solid angle
($\Ot$) of the torus as well as the luminosity of the corona (at energies
2--300~keV), $\Lc$. We can therefore use eq.~(\ref{eq:ltor}) to
study the relationship between $\Ld$ and $\Lc$, i.e., between 
emission properties of the accretion disk and corona. 

%%%%%%%%%%%%%%%%%%%%%%%%%%%%%%%%%%%
\subsection{Bolometric corrections}
\label{s:corrections}

We proceed by determining coefficients for conversion of the measured
quantities $\Lint$ and $\Lmir$ to the AGN intrinsic quantities $\Lc$
and $\Lt$, respectively. Hard X-ray spectral shapes do not vary much
from one Seyfert galaxy to another, apart from the 
photoabsorption rollover in type 2 objects below 10~keV. Typically,
absorption corrected AGN spectra can be described above 2~keV as a
power law with a photon index $\Gamma\sim 1.7$ (e.g., \citealt{reetur00})
and a rollover above $\sim 100$~keV (e.g., \citealt{moletal09}). We
adopt that  
\beq
\Lint\approx 0.3\Lc.
\label{eq:lint_lhx}
\eeq
This relation corresponds to a power-law spectrum with $\Gamma=1.7$ and an
exponential cutoff at $E_{\rm f}=200$~keV and is consistent with an 
average 3--300~keV spectrum of $\sim 100$ local AGNs detected 
during \integral\ and {\em RXTE} surveys of the sky
\citep{sazetal08b,sazetal10}. Assuming that the values of the  
power-law index and cutoff energy vary from $\Gamma=1.5$ to 1.9
and from $E_{\rm f}\sim 50$ to $\sim 500$~keV from one Seyfert galaxy
to another (as indicated by numerous studies, e.g.,
\citealt{zdzetal95,moletal09}), we can estimate that the conversion
described by eq.~(\ref{eq:lint_lhx}) can introduce a scatter in
$\Lint$ values for a given $\Lc$ of $\lesssim 20$\%, i.e., $\lesssim 0.1$~dex. 

We next introduce a similar correction factor for the reprocessed
emission from the torus:
\beq
\Lmir\approx 0.5\Lt.
\label{eq:lmir_ltor}
\eeq 

To obtain the above coefficient, we compared the 15~$\mu$m
luminosity with that integrated over the rest-frame 6--32~$\mu$m band,
$L_{6-32\,\mu\rm{m}}$, for those \spitzer/IRS spectra (21 in total)
that span this whole wavelength range (i.e., there are available data
from the IRS second-order SL module) and do not suffer from 
significant contamination by MIR emission from dust associated with star
formation. For the majority of these objects, the ratio
$\Lmir/L_{6-32\,\mu\rm{m}}$ is bounded in the narrow range 
of 0.65--0.85, only slightly depending on whether AGN emission lines
(such as [OIV] 25.9~$\mu$m) are taken into account or not. We
therefore adopted the relation $\Lmir/L_{6-32\,\mu\rm{m}}=0.75$ for AGN
tori and additionally lowered this ratio by one third in
eq.~(\ref{eq:lmir_ltor}) to account for non-negligible ($\sim 50$\%) additional
torus emission both shortward of 6~$\mu$m and longward of 
32~$\mu$m (see, e.g., \citealt{nenetal08}). While the
resulting $\Lmir/\Lt$ ratio (eq.~[\ref{eq:lmir_ltor}]) is determined
less strictly than the ratio $\Lint/\Lc$ above, the associated
scatter in individual $\Lmir/\Lt$ values around the mean trend
given by eq.~(\ref{eq:lmir_ltor}) is probably less than 20\%,
as suggested by the comparison of IRS spectra for ``pure'' objects,
described above.
 
%%%%%%%%%%%%%%%%%%%%%%%%%%%%%%%%%%%
\subsection{Solid angle of the torus}
\label{s:omega}

The next step in our analysis is to derive the torus solid angle
$\Ot$. To this end, we assume that, for a given hard X-ray
luminosity, $\Ot$ is determined by the relative number of
obscured (type~2) AGNs of that luminosity, i.e.,
\beq
\frac{\Ot}{4\pi}(\Lint)=
\frac{N_{\rm{type~2}}(\Lint)}{N_{\rm{total}}(\Lint)}.
\label{eq:omega}
\eeq
We consider an AGN obscured if its X-ray absorption column density
$\NH>10^{22}$~cm$^{-2}$. In this connection, recall (see \S\ref{s:sample}) 
that the $\NH$ values for our objects are not based on \integral\
hard X-ray measurements but have been determined through analysis of 
high-quality X-ray spectra obtained by various X-ray telescopes.

\begin{figure}
\centering
\includegraphics[bb=15 150 550 680,width=\columnwidth]{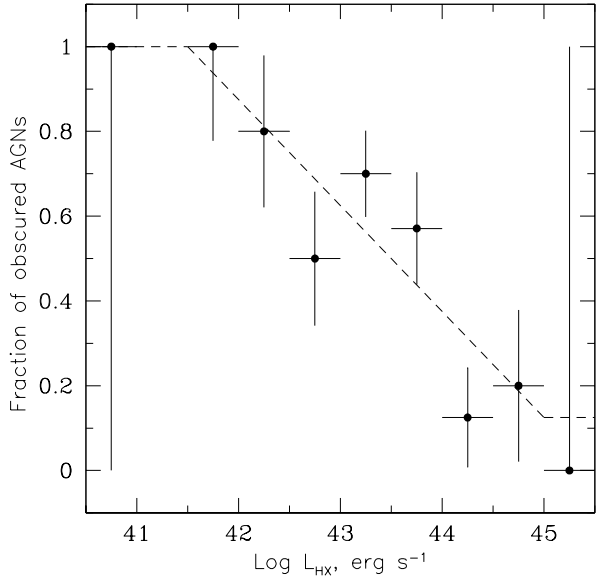}
\caption{Fraction of obscured ($\NH>10^{22}$~cm$^{-2}$) AGNs in the 
  local Universe as a function of hard X-ray luminosity, based on the
  \integral\ sample. The error bars represent the Poisson uncertainty associated
  with the number of objects in a given bin. The lowest and highest
  luminosity bins contain just one source each (NGC~4395 and IGR~J09446-2636,
  respectively). The approximate description of the observed trend by
  eq.~(\ref{eq:omega_lint}) is shown by the dashed line.
}
\label{fig:obsc_frac}
\end{figure}

Dividing our AGN sample into several bins in $\Lint$,
Fig.~\ref{fig:obsc_frac} shows the fraction of absorbed AGNs as a
function of $\Lint$. A strong trend of decreasing
$N_{\rm{type~2}}/N_{\rm{total}}$ ratio with increasing luminosity is 
evident. We can estimate the statistical significance of this 
trend using a maximum-likelihood estimator,
\beq
{\cal L}=-2\sum_{i}\ln P_{1,2} (\Linti).
\label{eq:like}
\eeq
Here, the summation is over our sample of AGNs and $P_{1,2}$ is the
probability for a given object with luminosity $\Linti$ to be either
obscured ($P_2$) or unobscured ($P_1$). We restrict  
our consideration to the luminosity range $41.5 < \log\Lint < 45$,
because there are only two objects in the sample which fall outside
this range (one on either side, see Fig.~\ref{fig:obsc_frac}). Suppose
now that the fraction of obscured AGNs has a power-law dependence on
luminosity:   
\beq
P_2=p+\alpha(\log\Lint-41.5).
\eeq
Obviously, 
\beq
P_1=1-P_2.
\eeq
Requiring that $0<P_{1,2}<1$ over the $41.5 < \log\Lint < 45$
range yields constraints on the slope and intercept of the trend:  
\beq
0<p<1
\label{eq:p}
\eeq
and 
\beq
-\frac{p}{3.5}<\alpha<\frac{1-p}{3.5}.
\label{eq:alpha}
\eeq
With these definitions, the maximum of the likelihood function proves
to be at $p\approx 1$ and $\alpha=0.25$. By integrating ${\cal
  L}$ over ($p$, $\alpha$) parameter space with the priors given
by eqs.~(\ref{eq:p}) and (\ref{eq:alpha}), i.e., using a Bayesian
approach, we find that the probability that $\alpha<0$ is
0.999. Hence, the declining luminosity trend of the obscured fraction
is ascertained with more than 3$\sigma$ significance.  

\citet{sazetal10} have recently confirmed this luminosity dependence
(see their Fig.~3) using a nearly doubled sample of AGNs detected
during 7 years of \integral\ observations (compared to the 3.5-year
all-sky survey used in the present work). Furthermore, the existence
of this trend has been reliably established in the past decade using
various X-ray selected samples of AGNs
(\citealt{uedetal03,steetal03,sazrev04,lafetal05,sazetal07,hasinger08,buretal11}; see in particular Fig.~8 in \citealt{hasinger08} and Fig.~13 in
\citealt{buretal11}). Therefore, in accordance with
eq.~(\ref{eq:omega}) we adopt the expression \beqa \frac{\Ot}{4\pi}
&=& \left\{
\begin{array}{l}
1,\,\,\,\log\Lint\le 41.5\\
1-0.25(\log\Lint-41.5),\\
\multicolumn{1}{r}{41.5<\log\Lint<45}\\
0.125,\,\,\,\log\Lint\ge 45.
\end{array}
\right.
\label{eq:omega_lint}
\eeqa

We thus assume that the phenomenon of decreasing fraction of absorbed 
AGNs with increasing luminosity reflects an underlying trend of
increasing opening angle of the obscuring torus. This is one of the
crucial points in our analysis. According to
eq.~(\ref{eq:omega_lint}), the slope of the $\Ot$ ($\Lint$) dependence
is approximately equal to 0.25 for $\Lint$ ranging between $\sim
10^{41.5}$ and $10^{45}$~erg~s$^{-1}$ 
with the associated uncertainty being small, $\sim 10$\% as determined 
from the dispersion of data points in Fig.~\ref{fig:obsc_frac} and
from the Bayesian analysis described above. However, 
the $\Ot$ ($\Lint$) dependence holds true in a statistical sense only,
and there might be variations in the torus
opening angle among AGNs of a given luminosity. Unfortunately,
observations do not yet provide reliable information on the
distribution of $\Ot$ values for a given $\Lint$, and hence we cannot
predict to what degree this scatter could affect our results below.
Furthermore, the exact parameters of the
$N_{\rm{type~2}}/N_{\rm{total}}$ ($\Lint$) dependence adopted in
eq.~(\ref{eq:omega_lint}) should be applied only to the local 
($z\sim 0$) population of AGNs, in particular because the
fraction of obscured sources among high-luminosity AGNs appears to be
larger in the distant ($z\gtrsim 1$) Universe (e.g.,
\citealt{hicetal07,hasinger08,treetal08}). 

%%%%%%%%%%%%%%%%%%%%%%%%%%%%%
\subsection{Disc vs. Corona}
\label{s:disk_corona}

\begin{figure}
\centering
\includegraphics[bb=15 150 550 680,width=\columnwidth]{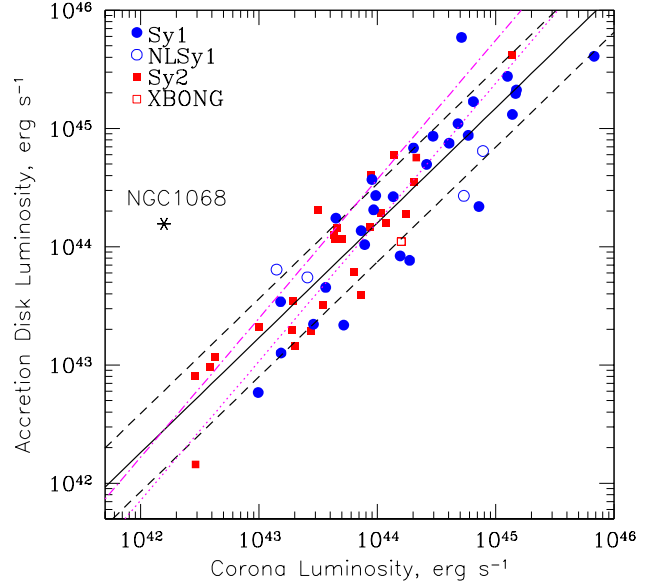}
\caption{Inferred luminosity of the accretion disk vs. that of the hot
  corona for the clean sample of 60 AGNs (the LLAGN NGC~4395 is excluded). 
  Various AGN types are indicated by different symbols as in
  Fig.~\ref{fig:xray_15m_lum_types}. Also shown is the Compton-thick
  Seyfert 2 galaxy NGC~1068, which was excluded from the analysis. The
  black solid line is the best-fitting power law dependence $\Ld$
  ($\Lc$) given by eq.~(\ref{eq:lc_ld}), while the two black dashed
  lines show this dependence multiplied and divided by 2.14, the rms
  scatter of the correlation. The magenta dotted line is the
  best-fitting dependence $\Lc$ ($\Ld$) (eq.~[\ref{eq:ld_lc}]),
  and the magenta dash-dotted line is the same dependence corrected
  for the Malmquist bias (eq.~[\ref{eq:ld_lc_malm}]).
}
\label{fig:corona_disk}
\end{figure}

We are now ready to estimate the accretion disk
luminosities for our AGNs using eq.~(\ref{eq:ltor}):
\beq
\Ld\approx \frac{4\pi}{\Ot(\Lint)} \Lt.
\label{eq:ld}
\eeq
Specifically, we first determine $\Lc$ and $\Lt$ using
eqs.~(\ref{eq:lint_lhx}) and (\ref{eq:lmir_ltor}), respectively and
then use eq.~(\ref{eq:ld}) to find $\Ld$. The resulting scatter plot
of $\Ld$ vs. $\Lc$  for the clean sample excluding NGC~4395 is shown
in Fig.~\ref{fig:corona_disk}.  

Fitting the $\Ld$ vs. $\Lc$ data in Fig.~\ref{fig:corona_disk} with a 
power law yields the correlation  
\beq
\Ldn=(1.59\pm0.16)\Lcn^{0.97\pm0.06},
\label{eq:lc_ld}
\eeq
where the luminosities are measured in units of $10^{44}$~erg~s$^{-1}$. 
The rms scatter around the mean trend is 0.33 and 0.34~dex
along the $\Ld$ and $\Lc$ coordinates, respectively. 

The derived relation, eq.~(\ref{eq:lc_ld}), allows one to predict the
disk luminosity for a given coronal luminosity. Hence, if the coronal
luminosity $\Lc$ of an AGN is known, one can expect its accretion disk
luminosity to be equal within a factor of $\approx 2$ (at the
1$\sigma$ confidence level) to $1.6\Lc$. The $\Ld/\Lc$ ratio does not
significantly depend on luminosity in the effective range of $\Lc$
from $\sim10^{43}$ to $10^{45}$~erg~s$^{-1}$.

The 2--10~keV energy band contains $\sim 25$\% of the total coronal
luminosity. If all of this X-ray emission were converted in 
the torus into infrared radiation as efficiently as accretion disk
emission, it would increase $\Lt$ by only $\sim 15$\%. This 
justifies the approximation adopted in eq.~(\ref{eq:ltor}). 
 
For some applications, one may also be interested in knowing the
distribution of $\Lc$ for a given $\Ld$, rather than $\Ld$ as a
function of $\Lc$. We have therefore also computed
(Appendix~\ref{s:app_inverse}) the inverse linear regression, i.e.,
$\log\Lc$ as a function of $\log\Ld$, for our clean sample
excluding NGC~4395. As can be seen in Fig.~\ref{fig:corona_disk},
this relation is different from the dependence of $\Ld$ on $\Lc$.

%%%%%%%%%%%%%%%%%%%%%%%%%%%%%%%%%%%%%%%%%%
\subsection{Scatter around the mean trend}
\label{s:scatter}

\begin{figure}
\centering
\includegraphics[bb=15 150 550 680,width=\columnwidth]{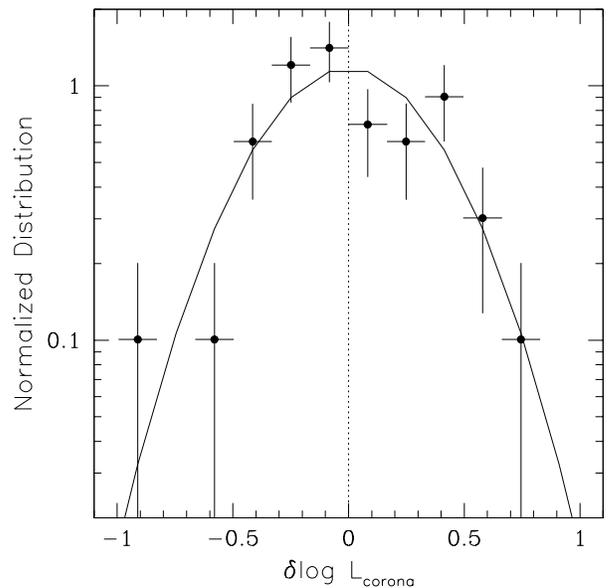}
\caption{Binned distribution of the residuals of the $\Lc$--$\Ld$ correlation, 
  eq.~(\ref{eq:lc_ld}). The error bars represent the Poisson uncertainty 
  associated with the number of objects in a given bin. The
  solid line shows the log-normal distribution corresponding to the
  measured rms scatter ($\sigma=0.34$).   
}
\label{fig:corona_disk_distr}
\end{figure}

Fig.~\ref{fig:corona_disk_distr} shows the distribution of residuals
for the $\Lc$--$\Ld$ correlation. Although it is plotted in terms of
$\Lc$ deviations, the distribution of
$\delta\log\Ld\equiv\log\Ld-\langle\log\Ld(\Lc)\rangle$ residuals is
quite similar. The distribution can be well-described by a log-normal function,
$dN/d\log\Lc\propto\exp[-(\delta\log\Lc)^2/2\sigma^2$], where
$\sigma=0.34$ is the measured rms scatter of the correlation.  

Although the observed scatter in the correlation is fairly small, the 
intrinsic correlation between the corona and disk luminosities in
Seyfert galaxies is probably even tighter. Indeed, the 
15~$\mu$m fluxes measured by \spitzer, from which the $\Ld$ values were
derived, presumably represent reprocessed accretion disk emission
averaged over a number of years. Interferometric observations of
Seyfert galaxies have established that the size of the MIR-emitting
dust region is of the order of several light years or more
\citep{trietal09}. Specifically, the characteristic size of the
12~$\mu$m source was found to scale approximately as the square root
of the AGN luminosity\footnote{This is consistent with the simple
  argument based on considering dust heating by a central source of UV
  emission \citep{barvainis87}.} and to vary from $l\sim 0.5$~pc for
$L_{12\,\mu\rm{m}}\sim 10^{42}$~erg~s$^{-1}$ to $l\sim 50$~pc for
$L_{12\,\mu\rm{m}}\sim 10^{46}$~erg~s$^{-1}$. For example,
\cite{trietal09} and \cite{trisch11} found $l\approx 2$--3~pc for
NGC~1068, NGC~1365, MCG~5-23-16, and NGC~4151 and $\sim 10$~pc for
NGC~7469. Therefore, the $\Ld$ values used in our cross-correlation
analysis should represent accretion disk luminosities averaged over
the $\sim 2l/c$-long period immediately preceding the
\spitzer\ observation of a given AGN, which is expected to
range from a few years for the least luminous sources to a few tens of
years for the most luminous ones. As demonstrated in
Appendix~\ref{s:app_var}, variability of the HX coronal emission,
detected by \integral\ on time scales shorter than the
characteristic time scale of MIR variations, is expected to induce a
scatter of $\sim 0.2$--0.25~dex around the mean $\Lc$--$\Ld$ trend. 

Additional contributions to the observed scatter around the mean
$\Lc$--$\Ld$ trend can be provided by systematic uncertainties
associated with: i) measurement of $\Lint$ and $\Lmir$, each
$\lesssim$~0.1~dex (see Table~\ref{tab:agn} and
\S\ref{s:fmir_uncert}), ii) conversion from $\Lint$ to $\Lc$ and from
$\Lmir$ to $\Lt$, also $\lesssim 0.1$~dex each
(\S\ref{s:corrections}), and iii) determination of the mean torus
solid angle ($\Ot$) for a given AGN luminosity and consequently
conversion from $\Lt$ to $\Ld$ (via eq.~[\ref{eq:ld}]), $\lesssim
  0.05$~dex (\S\ref{s:omega}). Hence, each of the above 
effects can contribute of the order of, or less than, 0.1~dex to the
observed scatter in the $\Lc$--$\Ld$ correlation. Adding these
contributions in quadrature to that expected from varibility implies
that the total induced scatter is $\sim 0.25$~dex. After subtraction
of this contribution from the measured scatter of the $\Lc$--$\Ld$
correlation, with rms $=0.33$~dex, there remains a scatter $\sim
0.2$--0.25~dex, i.e., a factor of 1.5--2. 

Therefore, the intrinsic correlation between accretion disk and
coronal emission in Seyfert galaxies is fairly tight. Furthermore,
neither the $\Lc$--$\Ld$ relation nor the $\Lint$--$\Lmir$ relation,
from which it originates, shows a significant dependence on either
optical AGN type or X-ray absorption column density, although there
are exceptions, which are discussed below.  Assuming that the torus is
co-aligned with the accretion disk and is a quasi-isotropic MIR
emitter, this suggests that the coronal hard X-ray emission is at most
modestly (less than a factor of $\sim 2$) anisotropic. This conclusion
also holds true if the obscuring torus is oriented quasi-randomly with
respect to the accretion disk/corona axis because for given $\Lmir$
there is little scatter in $\Lint$. By the same argument, the hard
X-ray luminosity of AGNs cannot be dominated by collimated emission
from relativistic jets.

As to the origin of the remaining (unaccounted for) scatter, at least
two effects are likely to contribute to it. First, our analysis was
based on the assumption that the characteristic solid angle subtended
by the obscuring torus, $\Ot$, is the same for all AGNs of a given
luminosity. In reality, it is possible that $\Ot$ varies from
one object to another (e.g., \citealt{elitzur12}). This would directly
affect our estimates of $\Ld$ from $\Lt$ and introduce scatter in the
resulting correlation between $\Lc$ and $\Ld$. Similarly, the
amplitude of the Compton reflection component, ignored in our
analysis, may vary from one AGN to another, which would introduce
additional scatter. Therefore, the scatter in the intrinsic ratio of
powers generated in the accretion disk and corona is likely even
smaller than 0.2~dex.

%%%%%%%%%%%%%%%%%%%%%%%%%%%%%%%%%%%%%%%%%%%%
\subsection{Comparison with typical quasars}
\label{s:quasars}

The results of this work pertain to nearby Seyfert galaxies, 
and it is interesting to put them into the broader context of the cosmic 
history of SMBH growth. To this end, we compare our findings with   
the properties of the SED of the ``average quasar'' from \cite{sazetal04}. 
This template essentially rests on the assumption that the cosmic 
X-ray background (CXB) represents the integrated hard X-ray emission of 
all AGNs in the Universe and on the argument of \cite{soltan82} that 
the cumulative bolometric luminosity of AGNs is determined by the mean 
radiative efficiency $\epsilon$ with which the integrated mass of local SMBHs 
has been accumulated over the cosmic time. Adopting $\epsilon=0.1$,
\cite{sazetal04} found that $\approx 12.5$\% of the bolometric
luminosity (below 300~keV) of the average quasar, is emitted at
energies above 2~keV. Attributing this emission to the corona of the
accretion disk and making a small correction for absorption in the
2--10~keV energy band (because here we are interested in intrinsic rather than
observed properties of accretion disks and coronae), we find that
$\Ld\approx 6\Lc$ for the average quasar. If we instead assume that 
$\epsilon=0.06$, which corresponds to the standard Shakura--Sunyaev
disk around a Schwarzschild black hole, then
$\langle\Ld/\Lc\rangle\approx 3.5$. By construction, this ratio
primarily characterizes quasars with $\Lc\sim 10^{44.5}$~erg~s$^{-1}$,
which produce the bulk of the CXB (e.g., \citealt{uedetal03}).  

We can now determine the corresponding average ratio for the local AGN
population, using a completely different method. The mean trend given
by eq.~(\ref{eq:lc_ld}) and the associated scatter (0.34~dex)
imply that the $\Ld/\Lc$ ratio varies between $\approx 0.75$ and
$\approx 3.4$ (the 1$\sigma$ range) around the mean value of $\approx
1.6$. Averaging over the log-normal distribution of $\Ld/\Lc$ yields
$\langle\Ld/\Lc\rangle=\int(\Ld/\Lc)\,d\Ld/\int\,d\Ld\approx
2.1$, independently of $\Lc$. This ratio characterizes the summed
radiation of the local AGN population. Therefore, the
$\langle\Ld/\Lc\rangle$ ratio appears to be larger, but only by a 
factor of $\sim 2$, for typical quasars making up the CXB relative to
typical AGNs in the local Universe. This implies that the {\sl ratio
  of the disk and coronal luminosities is approximately constant in
all actively accreting, radiatively efficient SMBHs.} 

%%%%%%%%%%%%%%%%%%%%%%%%%%%%%%%%%%%%%%%%%%%%%%%%%%%%%%%%%%%%%%%%%%%%%
\subsection{Possible effect of anisotropic accretion disk emission}
\label{s:cosine}

\begin{figure}
\centering
\includegraphics[bb=15 150 550 680,width=\columnwidth]{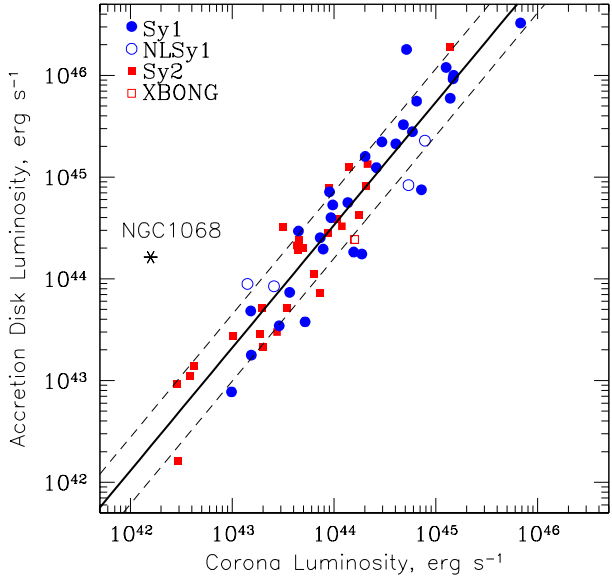}  
\caption{As Fig.~\ref{fig:corona_disk}, but for an accretion disk
  emitting according to the cosine law and a dusty torus lying in the
  plane of the disk. The best-fitting relation is given by
  eq~(\ref{eq:lc_ld_cos}).
}
\label{fig:corona_disk_cos}
\end{figure}

Our treatment so far has been based on the assumption that the
accretion disk is an isotropic source, i.e., its observed luminosity
is independent from the viewing angle. This implies that the primary
(UV) and reprocessed (infrared) luminosities of AGNs are related
through eq.~(\ref{eq:ltor}). However, the actual angular distribution
of radiation emergent from the accretion disk around a SMBH might be
close to Lambert's law, in which case the luminosity per solid
angle $d\Ld/d\Omega\propto\cos\theta$, where $\theta$  
is the viewing angle with respect to the axis of the
disk. If, in addition, an obscuring torus is not randomly oriented but
its equator lies in the plane of the accretion disk, then there will
be a different relation between $\Lt$ and $\Ld$:
\beq \Lt\approx\left(\frac{\Ot}{4\pi}\right)^2\Ld.
\label{eq:ltor_cos}
\eeq

Considering this possibility a feasible alternative to our baseline scenario, 
we repeated our calculations using eq.~(\ref{eq:ltor_cos}) instead of
eq.~(\ref{eq:ltor}). The resulting correlation between the corona and
disk luminosities is shown in Fig.~\ref{fig:corona_disk_cos} and the
corresponding best-fitting relation is given by:
\beq
\Ldn=(3.4\pm0.3)\Lcn^{1.21\pm0.06},
\label{eq:lc_ld_cos}
\eeq
with the rms scatter $\sigma=0.33$ in $\Ld$.

The anisotropic scenario predicts significantly larger $\Ld$/$\Lc$ ratios
at high luminosities (a factor of $\sim 3$ at $\Lc=10^{45}$~erg~s$^{-1}$) 
compared to the isotropic case. However, at these luminosities 
there is large uncertainty in the inferred $\Ld$ values
for the anisotropic case, which is not fully reflected in the formal
uncertainties quoted in eq.~(\ref{eq:lc_ld_cos}). Specifically, the 
uncertainty associated with our adopted dependence of the torus solid 
angle on luminosity, eq.~(\ref{eq:omega_lint}), is not taken into account. 
As is clear from Fig.~\ref{fig:obsc_frac}, this additional uncertainty is 
large at high luminosities, $\Lint\gtrsim 10^{44}$~erg~s$^{-1}$, which was   
relatively unimportant when we considered the isotropic scenario but
is the dominant source of uncertainty for the anisotropic
scenario due to the quadratic dependence on $\Ot$ in eq.~(\ref{eq:ltor_cos}).  

The above consideration ignores two potentially important
effects. First, radiative transfer in the accretion disk may  
lead to a limb-darkening effect, making the emission even more
collimated in the polar direction than $\cos\theta$ (e.g.,
\citealt{suntit85,laonet89}). On the other hand, strong gravity in the
vicinity of a black hole tends to harden the spectrum and increase
the luminosity for observers viewing the accretion disk at grazing
angles, with this effect being especially pronounced for rapidly
rotating black holes \citep{cunningham75,malkan83}.  

Taking these various factors into account, it is likely that the correlations
obtained in our isotropic (Fig.~\ref{fig:corona_disk}) and anisotropic
(Fig.~\ref{fig:corona_disk_cos}) scenarios bracket the true
relationship between the corona and disk emission in AGNs. Furthermore,
orientation effects and object-to-object variations in the black hole
spin can contribute to the scatter in the relation between $\Ld$ and
$\Lc$. 

%%%%%%%%%%%%%%%%%%%%%%%%%%%%%%%%%%%%
\section{Discussion and conclusions}
\label{s:discuss}
%%%%%%%%%%%%%%%%%%%%%

The main result of this work is that the luminosities of the accretion
disk and corona are nearly proportional for nearby AGNs:
$\Ld\propto\Lc^{0.97\pm0.06}$. To reveal this correlation, we derived
intrinsic (presumably emitted between $\lambda\sim 1$~$\mu$m and
$E\sim 2$~keV) accretion disk luminosities from observed torus
luminosities, $\Lt$, using eq.~(\ref{eq:ld}), which assumes that
radiation from the accretion disk is reprocessed in a dusty torus
whose effective solid angle decreases with increasing luminosity from
almost $4\pi$ at $\Lint\sim 10^{42}$~erg~s$^{-1}$ to $\sim 4\pi/8$ at $\Lint\sim
10^{45}$~erg~s$^{-1}$, as suggested by the observed falling fraction
of obscured AGNs (eqs.~[\ref{eq:omega}] and
[\ref{eq:omega_lint}]). This effect of decreasing obscuration fraction
is responsible for the MIR/HX luminosity ratio decreasing with   
increasing AGN luminosity: $\Lmir\propto\Lint^{0.74\pm0.06}$.

The observed $\Lc$--$\Ld$ relation implies a mean ratio
$\langle\Ld/\Lc\rangle\approx 2$ for nearby AGNs. For comparison, for
typical quasars producing the cosmic X-ray background,
$\langle\Ld/\Lc\rangle\sim 3.5$--6 (see \S\ref{s:quasars}). Hence, hard 
X-ray radiation from accretion disk coronae (with a possible
contribution from jets) carries a significant and roughly constant
fraction, $\sim 15$--35\%, of the bolometric luminosity of SMBHs
accreting in radiatively efficient mode (with a possible exception of
black holes accreting near the critical Eddington rate, see
\S\ref{s:edd} below).  

%%%%%%%%%%%%%%%%%%%%%%%%%%%%%%%%%%%%%%%%%%%%%%%%%%%%%%%%%%%%
\subsection{Intrinsic ratio of disk and corona luminosities}
\label{s:int_ratio}

The measured $\Ld/\Lc$ ratio likely overestimates the ratio of the
intrinsic UV and hard X-ray luminosities produced by the accretion
disk and corona, respectively, because roughly half of the luminosity
emitted by the corona is intercepted by the disk (and a small additional
amount by the obscuring torus) and only $\sim 10$--20\% of this
radiation (e.g., \citealt{haamar93}) is reflected, while the rest is
reprocessed into thermal, softer emission and thus contributes to the
disk luminosity. In contrast, it is well-known that the corona in
Seyfert galaxies intercepts only a small fraction of the disk's
radiation, probably because the corona is patchy (e.g.,
\citealt{zdzetal97}). Indeed, the measured shapes of the hard X-ray
spectra of Seyferts imply that the hot corona is characterized by an
amplification factor $A\sim5$--10 \citep{giletal00}, i.e.,
the luminosity of the Comptonized hard X-ray radiation emergent from
the corona is several times the luminosity of the incident soft
photons. This would imply that $\Ld\ll\Lc$ if the disk were embedded
in a homogeneous corona. Since in reality $\Ld\gtrsim\Lc$, as suggested by
the results of the present study, the presence of a strong big blue bump in
the spectra of type~1 AGNs, and theoretical arguments (e.g.,
\citealt{zdzetal97}), the solid angle subtended by the corona with respect
to the accretion disk must be small, i.e., to a first approximation
the corona does not shield the disk from the observer. 

On the other hand, the Compton reflection hump is located
approximately in our working \integral\ energy band (17--60~keV) and
may significantly contribute to the measured hard X-ray flux (see 
examples of tentatively detected reflection components in
\integral\ spectra of AGNs in \citealt{deretal12}), so that we likely
overestimate the flux of HX radiation coming
directly from the corona in our simplistic analysis. 

Considering these counteracting effects together, the intrinsic ratio
of the disk and coronal luminosities could be a factor of $\lesssim
2$ smaller than the measured $\Ld/\Lc$ ratio. {\em This implies that
  in Seyfert galaxies, approximately equal powers are generated in the
  accretion disk and hot corona.} 

%%%%%%%%%%%%%%%%%%%%%%%%%%%%%%%%%%%%%%%%%%%%%%%%%%%%%%%%%%%%%%%%%%%%
\subsection{AGN bolometric corrections. Current rate of SMBH growth}
\label{s:bolcorr}

Our results suggest that hard X-ray luminosity is a good proxy for
bolometric AGN luminosity ($\Lbol=\Ld+\Lc$), except for extremely
Compton-thick sources like NGC~1068. Using eqs.~(\ref{eq:lc_ld}) and
(\ref{eq:lint_lhx}), we can estimate the bolometric correction for the
17--60~keV energy band and the associated $1\sigma$ range (due to the
scatter in the $\Lc$--$\Ld$ correlation): 
\beq
\frac{\Lbol}{\Lint}\approx 9\,\,\,(6-15)
\label{eq:lbol_lint}
\eeq
in the range $\Lint\sim 10^{42}$--10$^{44.5}$~erg~s$^{-1}$. 

The cumulative hard X-ray (17-60~keV) luminosity density of nearby
AGNs found by integrating their luminosity function measured by
\integral\ at $\Lint\gtrsim 10^{41}$~erg~s$^{-1}$ is $(1.2\pm 0.2)\times
10^{39}$~erg~s$^{-1}$~Mpc$^{-3}$ \citep{sazetal07,sazetal10}. 
Low-luminosity ($\Lint\lesssim 10^{41}$~erg~s$^{-1}$) AGNs may add up
to $\sim 50$\% to this volume emissivity, as follows
from cross-correlating the cosmic X-ray background intensity with the
local galaxy distribution (e.g.,
\citealt{revetal08,miyetal94,caretal95}). Using the $\Lbol/\Lint$ 
ratio from eq.~(\ref{eq:lbol_lint}), the bolometric luminosity density
of local AGNs is thus $\sim (1-3)\times
10^{40}$~erg~s$^{-1}$~Mpc$^{-3}$. This implies that the integrated
present-day growth rate of SMBHs is $\dot{M}_{\rm SMBH}~(z=0)\sim
(2-5)\times 10^{-6}(0.1/\epsilon)~M_\odot$~yr$^{-1}$~Mpc$^{-3}$, where
$\epsilon$ is the average radiative efficiency of accretion. Comparing
$\dot{M}_{\rm SMBH}~(z=0)$ with the total mass density of local SMBHs,
$\rho_{\rm SMBH}~(z=0) \sim (3-5)\times 10^{5}M_\odot$~Mpc$^{-3}$ 
\citep{yutre02,maretal04}, the total SMBH mass is currently growing 
on a time scale $\sim 10$~times the age of the Universe. 

This estimate of the SMBH growth rate does not fully account for the
contribution of obscured accretion taking place in Compton-thick
($\NH\gg 10^{24}$~cm$^{-2}$) AGNs. Moreover, it is valid
only for accretion that is occurring in a radiatively efficient
mode. In reality, a substantial fraction of SMBH growth at the present 
epoch may be taking place through a radiatively inefficient mode of
accretion, dominated by mechanical rather than radiative energy output  
(e.g., \citealt{chuetal05,merhei08}). Therefore, the total SMBH
accretion rate may be higher.

We can re-calculate the bolometric correction given by
eq.~(\ref{eq:lbol_lint}) to the standard X-ray band,
2--10~keV. Assuming, as in \S\ref{s:corrections}, a power-law spectrum 
with $\Gamma=1.7$ and an exponential cutoff with $E_{\rm f}=200$~keV,
the 2--10~keV/17--60~keV luminosity ratio $\Lx/\Lint=0.82$ and
therefore
\beq
\frac{\Lbol}{\Lx}\approx 11\,\,\,(7-18).
\label{eq:lbol_lx}
\eeq
This formula predicts the bolometric luminosity of an
AGN from its intrinsic (i.e., unabsorbed) luminosity in the 2--10~keV
energy band in the range $\Lx\sim 10^{42}$--10$^{44.5}$~erg~s$^{-1}$.

Finally, we can estimate the bolometric correction for the MIR band
($\lambda=15~\mu$m), using eqs.~(\ref{eq:lbol_lint}) and
(\ref{eq:xray_15m_lum}):
\beq
\frac{\Lbol}{\Lmir}\approx 5.
\label{eq:lbol_lmir}
\eeq
This formula should be accurate to within a factor of $\sim 2$ for
AGNs with $\Lmir\sim 10^{42}$--10$^{44.5}$~erg~s$^{-1}$. We have not
tried to take into accout the non-linear dependence of $\Lmir$ on 
$\Lint$, because it should only be used to predict $\Lmir$ for a given
$\Lint$ but not $\Lint$ for a given $\Lmir$. A more reliable
bolometric correction for the MIR band could be obtained by using a
MIR selected sample of AGNs. Interestingly, eq.~(\ref{eq:lbol_lmir})
is in good agreement with an early estimate by \cite{spimal89},
$\Lbol/L_{12\,\mu\rm{m}}\sim 5$, based on direct integration of the
IR-to-UV ($\lambda=0.1$--100~$\mu$m) spectra of bright Seyfert
galaxies. In reality, as we have shown in this paper, the decreasing
trend of the MIR/bolometric luminosity ratio with increasing $\Lbol$
largely arises owing to the decreasing torus angular size, $\Ot$,
whereas the $\Ld/\Lbol$ fraction remains nearly constant. 

%%%%%%%%%%%%%%%%%%%%%%%%%%%%%%%%%%%%%%%%%%%%%%%%%%%%%%%%%%%%%%
\subsection{Comparison with the UV--X-ray luminosity relation}
\label{s:uv}

A number of studies have found that the ratio of near-UV ($\sim
2500\AA$) to soft X-ray ($\sim 2$~keV) luminosities in
type~1 AGNs increases with luminosity (e.g.,
\citealt{vigetal03,stretal05,steetal06,youetal10}; see, however,
\citealt{yuaetal98,gasetal04,tanetal07}). This suggests
that the disk/corona luminosity ratio increases with luminosity, in
apparent contradiction to our finding that this ratio is approximately
constant over about two decades in luminosity. 

Part of the explanation may be that, although the near-UV flux might
be a good proxy of the bolometric luminosity of the accretion disk in 
luminous quasars containing very massive black holes, it might be a poor
indicator (see also \citealt{vasfab07,vasetal09}) in lower luminosity
AGNs with less massive black holes (such as Seyfert galaxies), because
the maximum of their accretion disk emission is expected to be located
in the extreme-UV rather than in the near-UV band (e.g.,
\citealt{shasun73,hubetal01}). Mid-infrared observations, such as used
in the present study, make it possible to disclose the true bolometric
luminosity of the accretion disk by measuring the 
luminosity of the obscuring torus, which serves as a calorimeter of
the power radiated by the central engine.

Furthermore, there is probably no discrepancy at all, because our
study probes relatively low-luminosity AGNs compared to the quasars
used in the $\Lsx/\Lopt$ studies. Indeed, assuming again a
$\Gamma=1.7$ power-law spectrum with an exponential cutoff at $E_{\rm
  f}=200$~keV, the $10^{42}$--$10^{44.5}$~erg~s$^{-1}$ luminosity
range effectively probed by \integral\ in the 17--60~keV band
corresponds to a range of 2~keV-monochromatic luminosities of $9\times
10^{23}$ to $3\times 10^{26}$~erg~s~Hz$^{-1}$. As can be seen, e.g.,
from Fig.~4 in \citet{steetal06}, such AGNs are located in the
low-luminosity part of the $\Lopt$--$\Lsx$ diagram, where the data are
consistent with $\Lsx$ being proportional to $\Lopt$, whereas the 
evidence for a decreasing luminosity trend of $\Lsx$/$\Lopt$ comes
from more luminous AGNs with $\Lsx\gtrsim$~a few $\times
10^{26}$~erg~s$^{-1}$~Hz$^{-1}$. This possible change or gradual evolution of
the trend is further suggested by our finding (see \S\ref{s:quasars})
that for typical quasars producing the CXB, with $\Lsx\sim 2\times
10^{26}$~erg~s$^{-1}$~Hz$^{-1}$, the $\Ld/\Lc$ ratio is about twice that for
the \integral\ sample of (lower luminosity) AGNs. 

We conclude that although the reported behavior of the $\Lsx$/$\Lopt$
ratio may be indicative of the $\Ld/\Lc$ ratio decreasing with
luminosity in the most powerful quasars, this trend might be weak or
absent in less luminous AGNs. Clearly, further investigation of this
issue is required. 
 
%%%%%%%%%%%%%%%%%%%%%%%%%%%%%%%%%%%%%%%%%%%%%%%%%%%%%
\subsection{Implications for SMBH radiative feedback}
\label{s:feedback}

The results of the present work have implications for the role
of AGN feedback in the co-evolution of SMBHs and galaxies. 
The observed correlations between the masses of SMBHs and parameters
of their host elliptical galaxies/bulges (e.g., \citealt{tremaine02})
are possibly caused by coupling of energy released by the accreting
SMBH to gas inside and around its host galaxy. This energy can be supplied
in either mechanical (e.g., \citealt{chuetal02,king03}) or radiative
(e.g., \citealt{cioost01}) form. The latter possibility was considered
in detail by \cite{sazetal04}, who showed that radiation from typical
quasars producing the bulk of the CXB is characterized by a Compton
temperature\footnote{The Compton temperature is the temperature of a 
  gas interacting with a radition field at which there is no net
  energy exchange by Compton scattering between photons and
  electrons.} of $kT_{\rm c}\approx 2$~keV. From the
\integral/\spitzer\ study we now find, using eq.~[1] in
\citet{sazetal04} and taking into account the scatter in the $\Ld/\Lc$
ratio, that for local AGNs, $kT_{\rm c}$ varies between $\approx 2$
and 6~keV (note that the quasar SED in \citealt{sazetal04} extends to
MeV energies, while here we do not take into account any radiation
emitted above 300~keV).

Thus, radiation from a SMBH can photoionize and Compton heat ambient
interstellar gas above the virial temperatures of even giant elliptical
galaxies. Therefore, AGN radiative heating can indeed play an
important role in the co-evolution of galaxies and SMBHs, as has been
suggested previously \citep{cioost01,sazetal05,proetal08,novetal11}. As
discussed below, some studies suggest that the relative luminosity of
the corona becomes small for SMBHs accreting near the critical
Eddington rate, which would cause the Compton temperature to be
relatively low for such actively growing black holes. Therefore, there
is a need for a detailed study of the dependence of $T_{\rm c}$ on the
Eddington ratio.

%%%%%%%%%%%%%%%%%%%%%%%%%%%%%%%%%%%%%%%%%%%%%%%%%%%%%%%%%%%%%%%%%%%%%%%%%%%
\subsection{Dependence of the corona--disk relation on the Eddington ratio}
\label{s:edd}

\citet{vasfab07,vasfab09}, using UV and X-ray flux
measurements of AGNs and estimates of their black hole masses,
found that the disk/corona luminosity ratio is significantly larger
for SMBHs accreting close to the Eddington limit, i.e., having
$\Lbol/\Ledd\lesssim 1$, relative to objects with $\Lbol/\Ledd\ll
1$. However, this result is associated with significant uncertainty
(as emphasized by the authors), because the estimation of the
intrinsic (unabsorbed) luminosity of an accretion disk is difficult
even using UV data. 

It might be possible to reconcile the near constancy of the $\Ld/\Lc$
ratio found in our work (a similar conclusion also follows from the
study by \citealt {vasetal10} who analyzed {\em IRAS} infrared data for a {\em
  Swift}/BAT sample of AGNs) with the strong dependence of this ratio
on $\Lbol/\Ledd$ inferred by \citet{vasfab07} and \citet{vasfab09} if
we take into account the fact that most of the \integral\ (and {\em
  Swift}) AGNs have relatively low Eddington ratios
\citep{horetal12}. On the other hand, a few of our objects, in
particular the narrow-line Seyfert galaxies, have $\Lbol/\Ledd\lesssim
1$ but nevertheless occupy the same locus on the $\Lc$--$\Ld$ diagram
as the other objects (Fig.~\ref{fig:corona_disk}). This seems to 
argue against a strong dependence of $\Ld/\Lc$ on the Eddington
ratio. However, there is also a possiblity that the effective solid
angle of the obscuring torus explicitly depends on (decreases with)
the $\Lbol/\Ledd$ ratio, which would affect our estimates of the accretion disk
luminosity based on the infrared luminosity of the torus. 

In future work, an infrared--X-ray cross-correlation study based on
the \integral\ sample of local (relatively low $\Lbol/\Ledd$
ratio) Seyfert galaxies complemented by a representative sample of
high $\Lbol/\Ledd$ ratio quasars could help to clarify how the
disk/corona luminosity ratio depends on the Eddington ratio.

\begin{acknowledgments}
  This work is based on observations made with the \spitzer\ Space
  Telescope, operated by the Jet Propulsion Laboratory at Caltech
  under a contract with NASA, and with \integral, an ESA project funded
  by ESA member states (especially the PI countries: Denmark, France,
  Germany, Italy, Spain, Switzerland), Czech Republic and Poland, and
  with the participation of Russia and the USA. The research made use
  of data obtained through the High Energy Astrophysics Science
  Archive Research Center Online Service, provided by the NASA/Goddard
  Space Flight Center. The research made use of RFBR grants 09-02-00867 and
  11-02-12271-ofi-m. SS acknowledges the support of the Dynasty Foundation.
 
\end{acknowledgments}

%%%%%%%%%%%%%%%%
\begin{appendix}

%%%%%%%%%%%%%%%%%%%%%%%%%%%%%%%%%%%%%%%%%%%%%%
\section{IRS spectroscopy vs. IRAC photometry}
\label{s:app_irs_irac}
%%%%%%%%%%%%%%%%%%%%%%%%%%%%%%%%%%%%%%%%%%%%%%

To verify our conclusions about host galaxy contamination of IRS
spectra, we can compare 8~$\mu$m photometric measurements in small
(2.4$\arcsec$) and large (12$\arcsec$) IRAC apertures. The
result should depend on the extent of a starburst. If a star formation
region is more compact than the small (2.4$\arcsec$) photometric
aperture, there should be no significant associated extended 8~$\mu$m
flux. In the opposite case of a starburst extending over $\gtrsim
10\arcsec$, i.e., beyond both the large IRAC aperture and the IRS
extraction aperture, each 10\% starburst contribution to the AGN
spectrum at 15~$\mu$m should add $\sim 30$\% to the ratio of 8~$\mu$m
fluxes in the $12\arcsec$ and $2.4\arcsec$ apertures, as follows from
the comparison of our adopted starburst spectral template
\citep{braetal06} with a typical AGN spectrum (taking into account
that there are strong 7.7~$\mu$m and 8.6~$\mu$m PAH features falling
into the 8~$\mu$m IRAC filter). In addition, there might be a
non-negligible contribution of host galaxy starlight to the IRAC  
extended flux at 8~$\mu$m. In fact, extended stellar emission
is clearly seen by IRAC in most of our sources, including  ``pure'' AGNs,
at wavelengths $\lambda\lesssim 5$~$\mu$m (see
Fig.~\ref{fig:spectra}). The starlight contribution to the extended
flux at 8~$\mu$m can be roughly estimated as the extended 3.6~$\mu$m
flux multiplied by $(3.6/8)^2$ (i.e., assuming a Rayleigh--Jeans
spectrum).  

\begin{figure}
\centering
\includegraphics[bb=15 150 550 680, width=\columnwidth]{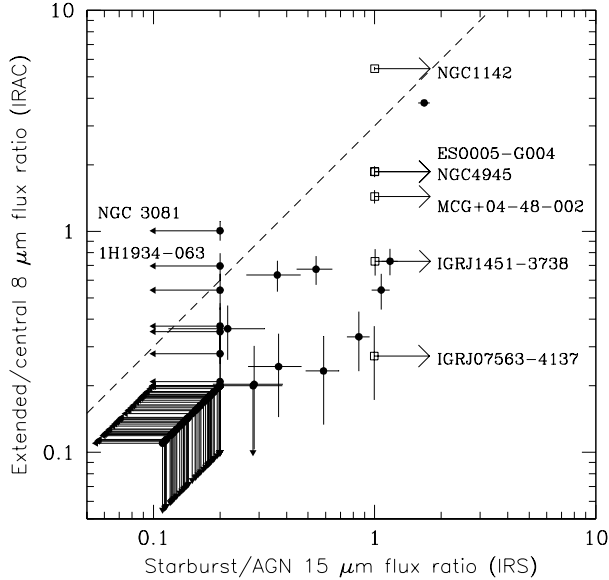} 
\caption{Starburst/AGN 15~$\mu$m flux ratio inferred from IRS spectra
  vs.  extended/compact 8~$\mu$m flux ratio determined by IRAC
  photometry in 12$\arcsec$ and 2.4$\arcsec$ apertures (with estimated
  starlight contribution subtracted from the extended flux). Filled
  circles denote objects from the clean sample (with AGN dominated 
  spectra). Non-detections ($<2\sigma$) in either of the two
  quantities are shown as upper limits of 20\% (shifted a little apart
  from each other in the lower left corner of the figure for better 
  visibility). Empty squares denote the 6 objects with starburst
  dominated IRS spectra, for which the starburst/AGN 15~$\mu$m flux
  ratio is of the order of 1:1 or larger. The dashed line indicates
  the $\sim 3:1$ extended/compact flux ratio expected for a sum of the
  adopted starburst template \citep{braetal06} and a typical AGN
  spectrum in the case of an extended star formation region.
}
\label{fig:irac_irs}
\end{figure}
 
Figure~\ref{fig:irac_irs} demonstrates that almost all of the sources
fulfil the above expectation, namely the extended/compact 
8~$\mu$m-flux ratio is less than $\sim 3$ times the starburst/AGN
spectral flux ratio at 15~$\mu$m. The positions of the sources
in the IRS--IRAC diagram probably reflect different angular sizes of
their starbursts: the further an object is located below the 3:1
limiting line the more compact is its star formation region. Two
sources, 1H~1934$-$063, and NGC~3081, lie somewhat above the 3:1
boundary in the IRS--IRAC diagram. This apparent
discrepancy probably arises because infrared spectral shapes of both
starbursts and AGNs should in fact differ from one object to another,
hence there may be a significant scatter around the $\sim 3:1$
relation between the starburst fractional contributions at 8 and
15~$\mu$m. We conclude that there is good overall agreement between
signatures of host galaxy contamination found by IRS spectroscopy and
by IRAC photometry for our objects.

%%%%%%%%%%%%%%%%%%%%%%%%%%%%%%%%%%%%%%%%%%%%%%%%%%%%
\section{Inverse correlations}
\label{s:app_inverse}
%%%%%%%%%%%%%%%%%%%%%%%%%%%%%%%%%%%%%%%%%%%%%%%%%%%%

The derived $\Lmir$ ($\Lint$) relation, eq.~(\ref{eq:xray_15m_lum}), 
makes it possible to predict the HX luminosity for a given MIR
luminosity. For some applications,
one may also be interested in knowing the distribution of $\Lint$ for
a given $\Lmir$. We have therefore also computed an inverse linear
regression, i.e., $\log\Lint$ as a function of $\log\Lmir$, for our
clean sample excluding NGC~4395:  
\beq
\Lintn=(0.81\pm0.13)\Lmirn^{0.98\pm0.08}.
\label{eq:15m_xray_lum}
\eeq
The rms scatter of $\Lint$ around the mean trend is 0.39~dex. 
As can be seen in Fig.~\ref{fig:xray_15m_lum}, this relation is
different from the dependence of $\Lmir$ on $\Lint$.

The dependence given by eq.~(\ref{eq:15m_xray_lum}) is
expected to be affected by the Malmquist bias. Indeed, our AGN sample
is hard X-ray selected. Therefore, if for a given MIR luminosity
$\Lmir$, there is a range of possible HX luminosities, $\Lint$, the
\integral\ survey would find more objects toward the higher boundary
of this range than toward its lower boundary because the more luminous
objects can be detected from larger distances and hence from a larger
volume of the Universe. For example, if for a fixed $\Lmir$, $\Lint$ varies from
object to object from $L_1$ to $L_2=4L_1$, then the survey will find
$4^{3/2}=8$ times as many $L_2$ sources as $L_1$ sources. The
intrinsic correlation can be found following \cite{viketal09} by
shifting the observed one (eq.~[\ref{eq:15m_xray_lum}]) by
$\Delta\log\Lint=-3/2\times\ln 10\times\sigma^2=-0.53$, where $\sigma=0.39$
is the measured rms scatter in $\log\Lint$. This results in 
\beq
\Lintn{\rm(corrected)}=(0.24\pm0.04)\Lmirn^{0.98\pm0.08}.
\label{eq:15m_xray_lum_malm}
\eeq

We can also compute the inverse relation between $\Ld$ and $\Lc$ for
our clean sample excluding NGC~4395: 
\beq
\Lcn=(0.67\pm0.06)\Ldn^{0.85\pm0.05}.
\label{eq:ld_lc}
\eeq 
It is also affected by the Malmquist bias. The intrinsic correlation
can be found by shifting the observed one (eq.~[\ref{eq:ld_lc}]) by
$\Delta\log\Lc=-3/2\times\ln 10\times\sigma^2=-0.33$ (where
$\sigma=0.31$ is the measured rms scatter in $\log\Lc$):  
\beq  
\Lcn {\rm(corrected)}=(0.31\pm0.03)\Ldn^{0.85\pm0.05}.
\label{eq:ld_lc_malm}
\eeq

Although the inverse relations, eqs.~(\ref{eq:15m_xray_lum_malm}) and
(\ref{eq:ld_lc_malm}), are formally correct, they should be used with
caution, because the implemented corrections for the Malmquist bias
are comparable to or larger than the intrinsic scatter in the
correlations (see Figs.~\ref{fig:xray_15m_lum} and
\ref{fig:corona_disk}). The inverse relations could be obtained more
reliably using a MIR-flux selected sample of AGNs. 

%%%%%%%%%%%%%%%%%%%%%%%%%%%%%%%%%%%%%%%%%%%%%%%%%%%%%%%%%%%%%%%%%
\section{Effect of X-ray variability on the derived correlations}
\label{s:app_var}
%%%%%%%%%%%%%%%%%%%%%%%%%%%%%%%%%%%%%%%%%%%%%%%%%%%%%%%%%%%%%%%%%

\begin{figure}
\centering
\includegraphics[bb=15 150 550 700,width=\columnwidth]{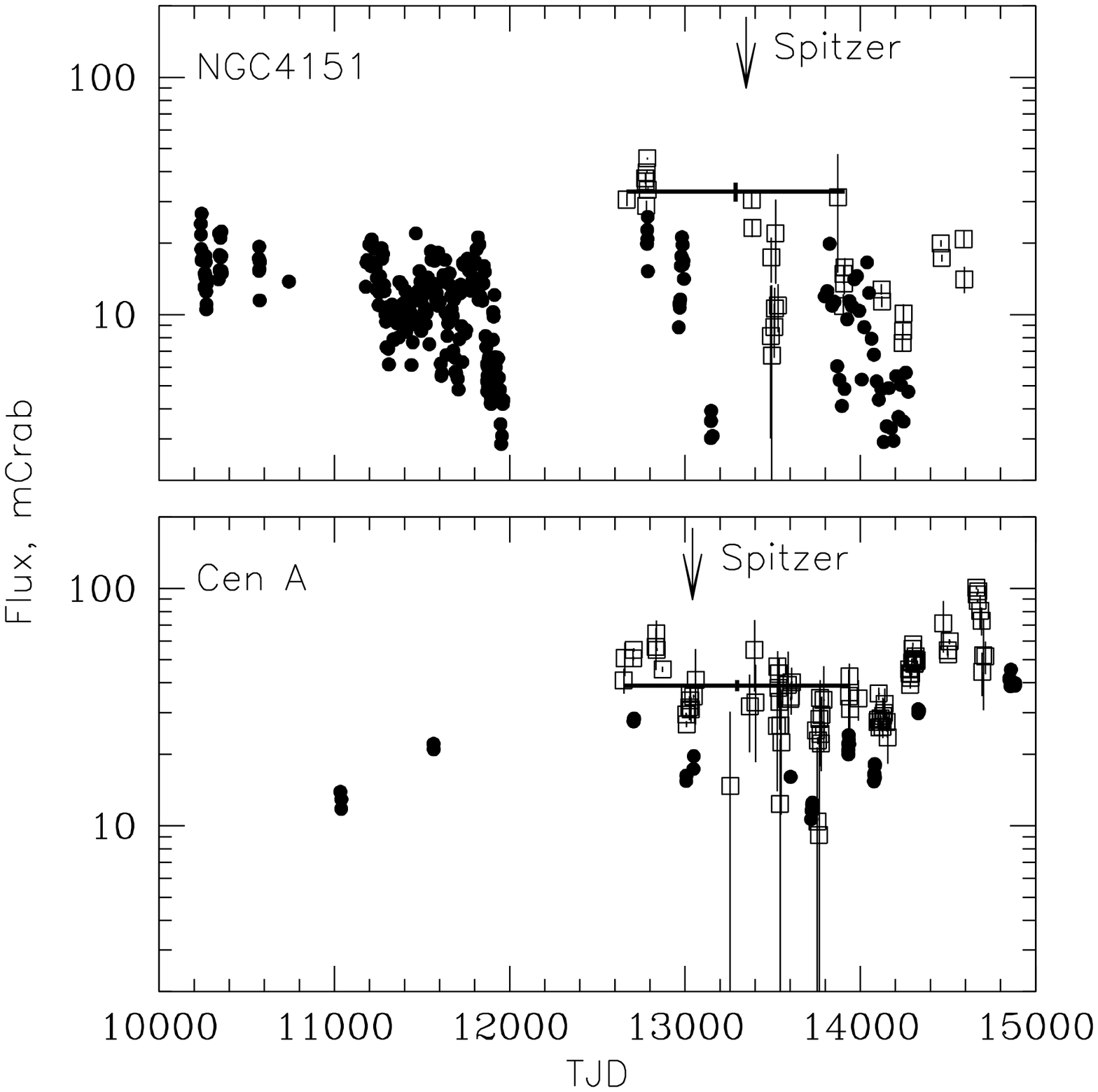}
\caption{Illustration of the effect of source variability on the
  $\Lc$--$\Ld$ correlation. Shown are the long-term light curves of
  NGC~4151 and Cen~A in the 3--20~keV ({\em RXTE}/PCA, filled circles)
  and 17--60~keV (\integral/IBIS, open squares) energy bands. The
  sampling time is 1 and 3~days for {\em RXTE} and \integral,
  respectively. Also indicated are the ``average'' \integral\ fluxes,
  $\fint$, and the periods used for their estimation (thick crosses),
  as well as the dates of the $\fmir$ measurements by
  \spitzer\ (arrows). Both $\fint$ and $\fmir$ have been used in the
  MIR--HX cross-correlation analysis. Since the MIR data represent the
  accretion disk emission averaged over the preceding period of a few
  years or longer, which does not coincide with the period over which
  the average HX flux was determined, it is clear that there should be
  significant associated scatter in the $\Lc$--$\Ld$ correlation.
}
\label{fig:lcurves}
\end{figure}

The coronal emission detected by \integral\ is expected to be
substantially variable on time scales much shorter than the
characteristic time scale of MIR variations, and this should affect
the observed $\Lc$--$\Ld$ correlation. Indeed, Seyfert galaxies are
known to be strongly variable in X-rays on time scales as short as
minutes and more so on times scales of months and years (see, e.g.,
\citealt{uttetal02}). To roughly estimate the effect of this
variability on the $\Lc$--$\Ld$ correlation, we have constructed
long-term (1996--2009) light curves in the 3--20~keV and 17--60~keV
energy bands of several bright, frequently observed AGNs from our
sample using {\em RXTE}/PCA and \integral/IBIS data, respectively.

\begin{figure}
\centering
\includegraphics[bb=15 150 550 700,width=\columnwidth]{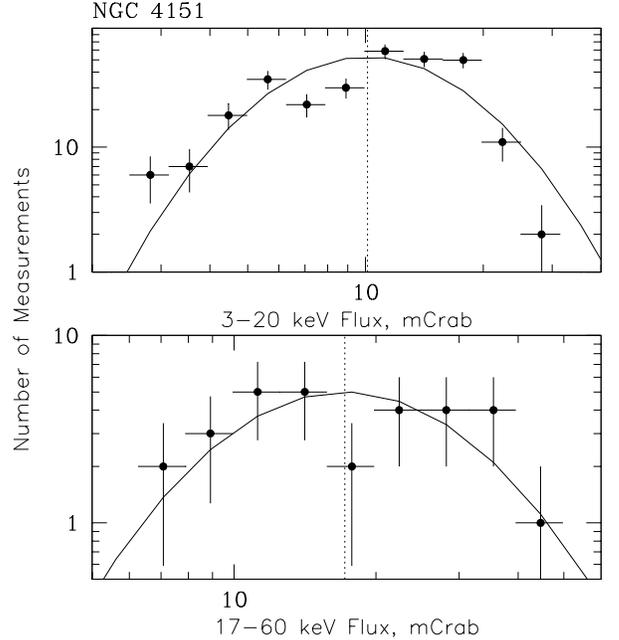}
\caption{Binned distribution around the log-mean value of individual
  flux measurements for NGC~4151 by {\em RXTE}/PCA in the
  3--20~keV energy band (upper panel) and by \integral/IBIS in the
  17--60~keV band (lower panel). The solid line shows the log-normal
  distribution with a standard deviation equal to the actually
  measured rms scatter, 0.22~dex in the 3--20~keV band and 0.24 in the
  17--60 keV band.
}
\label{fig:NGC4151_flux_distr}
\end{figure}

\begin{figure}
\centering
\includegraphics[bb=15 150 550 700,width=\columnwidth]{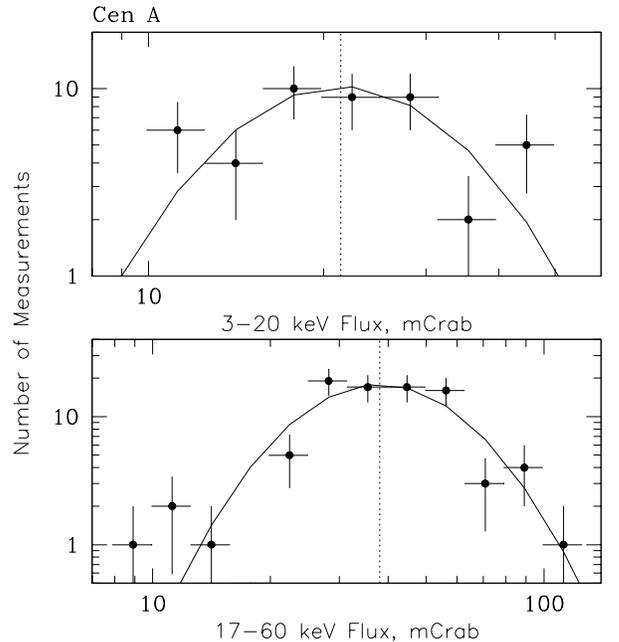}
\caption{As Fig.~\ref{fig:NGC4151_flux_distr}, but for Cen~A. The
  rms scatter around the mean $\log F$ is 0.17 and 0.19~dex in the
  3--20~keV and 17--60~keV energy bands, respectively.
}
\label{fig:CENA_flux_distr}
\end{figure}

Figure~\ref{fig:lcurves} shows the light curves of NGC~4151 and
Cen~A, the two brightest AGNs in our sample. The (longer and better
sampled) {\em RXTE} light curves demonstrate that the 3--20~keV fluxes
of both sources varied by an order of magnitude over a time span of a
decade. The same variability behavior is confirmed by the \integral\
data taken since 2002. The X-ray and hard X-ray variations are
quantified in Figs.~\ref{fig:NGC4151_flux_distr} and
\ref{fig:CENA_flux_distr}, which show the distribution of individual
flux measurements, binned in $\log$ of flux. The observed
flux distributions for NGC~4151 and Cen~A can be well described as
log-normal with $\sigma=0.22$--0.24 and 0.17--0.19~dex,
respectively. This result is consistent with a number of previous
studies that have demonstrated that X-ray flux variations in AGNs,
similarly to X-ray binaries, can be described by a log-normal
distribution \citep{gaskell04,uttetal05}.
 
The majority of AGNs in our sample were observed by \integral\ only
occasionally, typically once a year for a duration of a few days, hence
the HX fluxes measured during 2002--2006, which are used in our
cross-correlation analysis, in fact represent random snapshots of
sources rather than their long-term averaged fluxes. Therefore, if
NGC~4151 and Cen~A are typical of the whole sample, X-ray variability
should induce a scatter of $\sim 0.2$--0.25~dex around the mean
$\Lc$--$\Ld$ trend. To further illustrate this point, HX fluxes
averaged over a more recent period of \integral\ observations,
from 2006--2009, give a factor of $\sim 2$ smaller $\Lc$ for NGC~4151 and a
factor of $\sim 1.5$ larger $\Lc$ for Cen~A. 

\end{appendix}

%%%%%%%%%%%%%%%%%%%%%%%%%

\end{document}